\documentclass[pre,amsmath,amsfonts,amssymb,twocolumn]{revtex4}
\usepackage{amssymb}
\usepackage{bbm}
\usepackage{bm}
\usepackage{graphicx}
\usepackage{color}
\usepackage{dsfont}
\usepackage{amsmath}
\graphicspath{{./},{../../Mathematica/KIC/picdump/}}
\definecolor{orange}{RGB}{255,127,0}

\begin{document}
\title{Transition from Quantum Chaos to Localization in Spin Chains}

\author{Petr Braun$^1$, Daniel Waltner$^1$, Maram Akila$^1$, Boris Gutkin$^2$, Thomas Guhr$^1$}
\affiliation{$^1$Fakult\"at f\"ur Physik, Universit\"at Duisburg-Essen,
  Lotharstra\ss e 1, 47048 Duisburg, Germany\\
  $^2$Department of Applied Mathematics, Holon Institute of Technology, 58102 Holon,
Israel}

\begin{abstract}Recent years have seen an increasing interest in quantum chaos and related aspects of
spatially extended systems,
such as spin chains. However, the results are strongly system dependent,
generic approaches suggest the presence of many-body localization
while analytical calculations for certain system classes, here referred
 to as the ``self-dual case'', prove adherence to universal (chaotic)
spectral behavior. We address these issues studying the level statistics in the vicinity of the
latter case, thereby revealing transitions to  many-body localization as
well as the appearance of several non-standard random-matrix universality classes.
%Spin chains are quite topical systems in current research. In Ref.\ [M.\ Akila, D.\ Waltner, B.\ Gutkin, T.\ Guhr, J.\ Phys.\ A {\bf 49}, 375101 (2016)] we identified a subspace
%of the parameter space of the system where the change of {\it particle number} and {\it time} can be considered to be largely equilvalent termed the self-dual case. Here we extend previous studies on
%the spectral properties [B.\ Bertini, P.\ Kos, T.\ Prosen, Phys.\ Rev.\ Lett.\ {\bf121}, 264101 (2018)] valid in the self-dual case to its vicinity. In this context we reveal interesting
%symmetry properties of the spectrum that can be explained by new ensembles of Random Matrix Theory.
\end{abstract}

\maketitle

{\section{Introduction}}  An early root of quantum chaos \cite{Haake,Stoeckmann,Gutzwiller,Gnutz} was the study of  the level statistics in typical
many-body systems like nuclei.
Attempting to relate quantum to classical chaos, the complexity of the many-body dynamics brought single-particle systems into the focus.
In recent years,
the interest turned back to many-body systems \cite{Engl,Dubertrand,Michl,HammerlingII,Gessner,Tom,Tom1}. Ironically, although the dynamics in such systems is rather
complicated, it is not always chaotic, examples are  integrable dynamics, various types of regular collective motion \cite{Baldwin,Bohle} and many-body
localization (MBL) \cite{Altshuler,Basko,Znidaric,Moess,Schreiber,Choi}.

We consider spin chains that presently attract considerable interest
%\cite{Simon,Smith,Islam,Murmann,Kim,WiesendI,WiesendII,Akila0,Akila,Prosen0,Prosen,Prosen100,Chalker,Braun,Linden,Atas,Akila1}.
as they can be realized experimentally with cold atoms \cite{Neill,Simon,Smith,Murmann,Kim,Islam} or as chains on surfaces \cite{WiesendI,WiesendII}.
On the theoretical side various different directions in many-body chaos converge in spin chains, such as thermalization in finite systems \cite{Steinigeweg,Howell,Santos,Luitz2,Borg}, localization and entanglement effects
\cite{Ponte,Prosen100,Lak,Luitz,Luitz1,Gart,Barda,Wei}, spectral properties
\cite{Gharibyan,Chalker,Akila,Prosen0,Prosen,Linden,Atas,Schiu,Chalker1,Serbyn},
characterized e.g.\
by the spectral form factor, and the classical-quantum transition \cite{Akila0,Akila1,Braun,Waltner,Obertha,Gessner}.

We are interested in chains of $N$ spin-$1/2$ particles for arbitrary $N$.
%\cite{Neill,Simon,Smith,Islam,Murmann,Kim,WiesendI,WiesendII,Akila,
%Prosen0,Prosen,Linden,Atas,Gharibyan,Luitz,Luitz1,Gart,Obertha,Schiu,Barda,Wei,Serbyn,Steinigeweg,Howell,Santos,Borg}.
We employ duality: the unitary propagator in time corresponds to a nonunitary ``propagation'' in particle
number governed by an operator dual to the time evolution operator \cite{Akila,Bori}. By this we
calculate the
spectral form factor that
characterizes correlations between the eigenvalues of a system, as  a function of time in the disorder free case.
The Prosen group \cite{Prosen0} extended this including  disorder and confirmed the predictions made by Random Matrix Theory (RMT) \cite{Guhr,Metha} for long times.
The dual operator can be unitary \cite{Akila}, this situation was coined ``self-dual case'' \cite{Prosen}.
\textit{In the self-dual case} the exact RMT result for the spectral form factor applies at all fixed times in the thermodynamic limit of infinitely
long disordered chains \cite{Prosen}.

Here, we establish the symmetry property of the  kicked Ising chain similar to the charge conjugation \cite{Haake}; as a result at the point of self-duality its spectrum is described by
one of the "new" ensembles of RMT rather than the expected circular orthogonal ensemble (COE). Then we study the stability of the RMT results {\it as we go away from self-duality} addressing the questions:
How does the transition occur between RMT and the localized regime? Specifically, how does the spectral form factor change
for short times if we move away from the self-dual case? How does it change in dependence of time revealing ergodic behavior or localization in the long-time limit?
Such an investigation is all the more
important since a variety of seemingly close models of the disordered spin-$1/2$ chains exhibit qualitatively different spectral properties obeying RMT
\cite{Prosen,Prosen0} or indicating MBL
\cite{Chalker}.

We derive an exact analytical expression for the form factor for long chains near the self-dual case. Larger violation of self-duality brings about an explosive growth of the
spectral form factor that becomes sharp in the limit $N\to \infty$ similar to a thermodynamic phase transition leading finally to MBL.
We want to emphasize that these results go beyond the presented spin-$1/2$ model
as they hold for self-dual  models in general.

The outline of this paper is as follows: in the next section
we introduce the model and the dual operator and in Sec. III  the underlying
symmetries of the time evolution operator and the dual operator. In Sec. IV we
establish the spectral form factor. The functional form of the form factor is analyzed in Sec.
V and in Sec. VI we address the transition to localization in that system. In Sec. VII we conclude.

\section{Kicked Spin Chain}  We study a periodically kicked disordered chain
of $N$ spins with the Pauli matrices $\hat{\pmb{\sigma}}_n=\left(\hat{\sigma}_n^x,\hat{\sigma}_n^y,\hat{\sigma}_n^z\right)$, the  Ising coupling $J$, the magnetic field $b_x$
in $x$ direction and the site dependent field $h_{n}$ in $z$ direction modeling disorder \cite{Akila,Akila0,Akila1,Prosen100,Prosen}. The time evolution operator per period is
$\hat U=\hat{U}_I\hat{U}_b$, where
\begin{eqnarray}\label{timeevo}
\hat{U}_I&=&\exp\left(  -iJ\sum_{n=1}^{N}\hat{\sigma}_{n}^{z}\hat{\sigma}_{n+1}%
^{z}\right)\nonumber\\\hat{U}_b&=&\exp\left(-i\sum_{n=1}^{N}h_{n}\hat{\sigma}_{n}^{z}\right)  \exp\left(  -ib_{x}%
\sum_{n=1}^{N}\hat{\sigma}_{n}^{x}\right)
\end{eqnarray}
are
the Ising and the magnetic field part, respectively. In the $\hat{\sigma}_n^z$ product eigenbasis
 $\left|\vec{\sigma}^N\right\rangle=\left|\sigma_1,\ldots,\sigma_N\right\rangle$, its dimension is $2^N\times2^N$.

The in general nonunitary dual operator $\hat{W}$ of dimension $2^T\times2^T$ in the basis $\left|\vec{\sigma}^T\right\rangle=\left|\sigma_1,\ldots,\sigma_T\right\rangle$ is especially
suited for the small $T$ regime. It fulfills ${\rm{Tr}}\;\hat{U}^T={\rm Tr}\;\hat{W}$ with $\hat{W}=\prod_{n=1}^N\hat{W}_n$. Here $\hat{W}_{n}=\hat W_{I}\hat W_{n,b}$ with
\begin{eqnarray}\label{dualop}
\langle\vec{\sigma}^{T}{'}|\hat W_{I}|\vec{\sigma}^T\rangle
&=&\exp\left(  -iJ\sum_{t=1}^{T}{\sigma}_{t}{\sigma}_{t}'\right)  ,\\
\langle\vec{\sigma}^{T}{'} |\hat W_{n,b}|\vec{\sigma}^T\rangle
&=&\delta_{\vec{\sigma}\vec{\sigma}^{\prime}}\exp\left(  -ih_{n}\sum_{t=1}%
^{T}{\sigma}_{t}\right)  \prod_{t=1}^{T}R_{\sigma_{t}\sigma_{t+1}},\nonumber
\end{eqnarray}
in the basis $\left|\vec{\sigma}^T\right\rangle$, and  $R_{\sigma_{t}\sigma_{t+1}}$: $R_{11}=R_{-1-1}=\cos b_x,\hspace*{3mm}R_{-11}=R_{1-1}=-i\sin b_x$.
The  condition for self duality, i.e.\ for the unitarity of $\hat{W}$, is $J=b_x=\pi/4$.
This is so because $\prod_{t=1}^{T}R_{\sigma_{t}\sigma_{t+1}}$ then transforms into $(-1)^{\nu}2^{-T/2}$ with $2\nu$ being the number of domain walls
in the dual ring $\sigma_1,\ldots,\sigma_T$ leading, up to the factor $2^{-T/2}$, to a unitary operator $\hat W_{n,b}$.
Including this factor into $\hat W_{I}$ transforms this operator into the $T$ dimensional (unitary) discrete Fourier transform.

{\section{Symmetry Relations}}
\subsection{The evolution operator}
  The operator $\hat U$ admits time reversal, i. e., there exists an anti-unitary operator $\hat T$ such that $\hat T \hat U \hat T^{-1}=\hat U^{-1}$ \cite{Haake}; explicitly $\hat T= \exp\left(  ib_{x}%
\sum_{n=1}^{N}\hat{\sigma}_{n}^{x}\right)\, \hat K$ where $\hat K$ stands for complex conjugation in the standard basis. For large enough $N$ and random local fields $h_n$ one would expect that the spectrum of  $\hat U$ follows the COE predictions, with the eigenphase density constant on the unit circle.

 In fact, for a self-dual chain the symmetry of the problem is higher. To show it we start with the identity for the Floquet operator in the kicked Ising model which, to our knowledge, has not been previously reported,
\begin{eqnarray}
\hat{U}'=\left(-i\right)^N\hat{\Sigma}_y^N\hat{U}^*\hat{\Sigma}_y^N,\quad \hat{\Sigma}_y^{N}=\otimes_{n=1}^{N}\hat{\sigma}_n^y,\label{symmrelU}
\end{eqnarray}
where  the unprimed operator is evaluated at $J=\pi/4-\Delta J$ and the primed one at $J=\pi/4+\Delta J$; the fields $h_n$ and $b_x$ are arbitrary.

To prove Eq.\ (\ref{symmrelU}) note that $\hat{U}_b$ remains invariant under the simultaneous
conjugation with $\hat{\Sigma}_y^N$ and complex conjugation. The exponent $\sum_{n=1}^N\hat{\sigma}_n^z\hat{\sigma}_{n+1}^z$ of the Ising part $\hat{U}_I$  is diagonal in the $\left|\vec{\sigma}^N\right\rangle$-basis
with the value $(N-2\nu)(+1)+2\nu(-1)=N-4\nu$ with $2\nu$ being the (even) number of domain walls.  It follows that in Eq.\ (\ref{symmrelU}) the diagonal elements
of $\hat{U}_I^*$ are  $(-1)^{\nu}\exp\left[i(N-4\nu)\Delta J\right]{\rm e}^{iN\pi/4}$ whereas for $\hat{U}_I'$ they are $(-1)^{\nu}\exp\left[i(N-4\nu)\Delta
J\right]{\rm e}^{-iN\pi/4}$. This proves relation (\ref{symmrelU}).

The identity (\ref{symmrelU}) has important consequences for the spectral properties of $\hat{U}$  in the self-dual case when  $\Delta J=0$ such that the primed and non-primed operators coincide.
They are conveniently formulated for the operator,
$\hat U_M\equiv\exp\left(iN\pi/4\right)\hat{U}$. Then we have the symmetry relation,
\begin{eqnarray}\label{symmrelM}
\hat U_M= \hat{\Sigma}_y^{N}\hat U_M^* \hat{\Sigma}_y^{N}.
\end{eqnarray}
It can be rewritten as the commutation relations,
\begin{eqnarray}\label {commrel}
\left[\hat U_M,\hat C^N\right]=0,
\end{eqnarray}
where    $\hat C^N$ is an anti-unitary ``charge conjugation'' operator \cite{Haake}  defined like
\begin{eqnarray}
\label{charconj}
\hat C^{N}=\hat{\Sigma}_y^{N}\hat K,\qquad \hat C^2=(-1)^{N}.
\end{eqnarray}
In view of the unitarity of  $\hat U_M$  it follows that its
eigenphases come in complex conjugated pairs which singles out the points
$0$ and $\pm\pi$ in the spectrum.  After an appropriate transformation of the basis set its matrix becomes symplectic for $N$  odd
and orthogonal  for $N$ even.

To show this, consider  the relation (\ref{symmrelM}).
The operator $\hat{\Sigma}_{y}^{N}$ has non-zero matrix elements only between the
basis states $\left\vert \vec{\sigma}\right\rangle =\left\vert \sigma
_{1},\ldots,\hat{\sigma}_{N}\right\rangle $ and $\left\vert -\vec{\sigma
}\right\rangle ,$ i.e., with all spins flipped. Let $\eta\left(  \vec{\sigma
}\right)  =1$ if the number of spins-up in $\left\vert \vec{\sigma
}\right\rangle $ is even and $-1$ otherwise; obviously $\eta\left(
-\vec{\sigma}\right)  =\left(  -1\right)  ^{N}\eta\left(  \vec{\sigma}\right)
$. Then we have,
\begin{equation}\label{sigmamatrelt}
\left\langle \vec{\sigma}\left\vert \hat{\Sigma}_{y}^{N}\right\vert \vec{\sigma
}^{\prime}\right\rangle
 =\left(  -1\right)  ^{N}\left\langle \vec{\sigma}^\prime
\left\vert \hat{\Sigma}_{y}^{N}\right\vert \vec{\sigma}\right\rangle
=i^{N}\eta\left(  \vec{\sigma}\right)  \delta_{\vec{\sigma},-\vec{\sigma
}^{\prime}}.
\end{equation}
Let us group the basis states into pairs $\left\vert \vec{\sigma}\right\rangle
,\left\vert -\vec{\sigma}\right\rangle $ . First consider  $N$  odd and
choose  $\left\vert \vec{\sigma}\right\rangle $ in each pair such that
$\eta\left(  \vec{\sigma}\right)  =+1$. In that basis set  $\hat{\Sigma}_{y}%
^{N}=i^{N}\hat{\Omega}$ where $\hat{\Omega}$ is the fundamental symplectic matrix for the $q_1,p_1,\ldots,q_{2^{N-1}},p_{2^{N-1}}$-ordering of the canonical variables; it is composed of $2^{N-1}$
blocks $%
\begin{pmatrix}
0 & 1\\
-1 & 0
\end{pmatrix}
$ on its diagonal. Using unitarity of $\hat U_M$ we can rewrite the symmetry relation for it as%
\begin{equation}
\hat{U}_{M}^{-1}=\hat{\Omega}\,\hat{ U}_{M}^\mathcal{T} \,\hat{\Omega}.
\end{equation}
where $\mathcal{T} $ stands for transposition. By definition, it follows that $\hat{U}_{M}$ is symplectic.

If $N$ is even both members of the pair $\left\vert \vec{\sigma}\right\rangle
,\left\vert -\vec{\sigma}\right\rangle $ have the same $\eta$. The matrix
$\hat{\Sigma}_{y}^{N}$ is again block-diagonal, however its $2\times 2$ blocks are
now $i\,\eta\left(  \vec{\sigma}\right)  %
\begin{pmatrix}
0 & 1\\
1 & 0
\end{pmatrix}.
$
 Let us transform the basis set introducing, for each pair,%
\begin{align}%
\begin{pmatrix}
u_{\vec{\sigma}}\\
v_{\vec{\sigma}}%
\end{pmatrix}
& =\hat{q}\left(  \vec{\sigma}\right)
\begin{pmatrix}
\left\vert \vec{\sigma}\right\rangle \\
\left\vert -\vec{\sigma}\right\rangle
\end{pmatrix}
,\\
\hat{q}\left(  \vec{\sigma}\right)    & =%
\begin{pmatrix}
1 & -i\\
\eta\left(  \vec{\sigma}\right)   & -i\eta\left(  \vec{\sigma}\right)
\end{pmatrix}
.
\end{align}
Denoting by tildas matrices transformed to the $uv$-basis and introducing the transformation matrix $
\hat Q=\mathrm{diag} \{\ldots,\hat{q}\left(  \vec{\sigma}\right) ,\ldots\}$ we have
\begin{eqnarray}\label{trafomatQ}
\hat U_M=\hat Q^* \tilde U_M \hat Q^\mathcal {T}, \quad\hat{\Sigma}_{y}^{N}=\hat Q^* \tilde{\Sigma}_{y}^{N}\hat  Q^\mathcal {T}.
\end{eqnarray}
Considering that $\hat q^\mathcal{T}\hat q= \hat q^\dagger \hat q^*=\hat \sigma_z$ it follows that
\begin{eqnarray}
\tilde{\Sigma}_{y}^{N}=i^N \hat Z_{N}, \quad  \hat Q^\mathcal {T} \hat Q=\hat Q^\dagger \hat Q^*=\hat Z_{N}.
\end{eqnarray}
where $\hat Z_N$ stands for the $2^N\times 2^N$ diagonal matrix $\mathrm{diag}\{1,-1,\ldots,1,-1\}$. Substituting (\ref{trafomatQ}) into (\ref{symmrelM}) we obtain
the symmetry condition in the $uv$-basis as
$
\tilde U_M =(-1)^N\tilde U_M^*;
$
since $N$ is even $\tilde U_M$ is real unitary, i. e., orthogonal.

\begin{figure}
 \begin{center}
% \vspace*{-1cm}

  \hspace{-0.5cm}
\includegraphics[width=9cm]{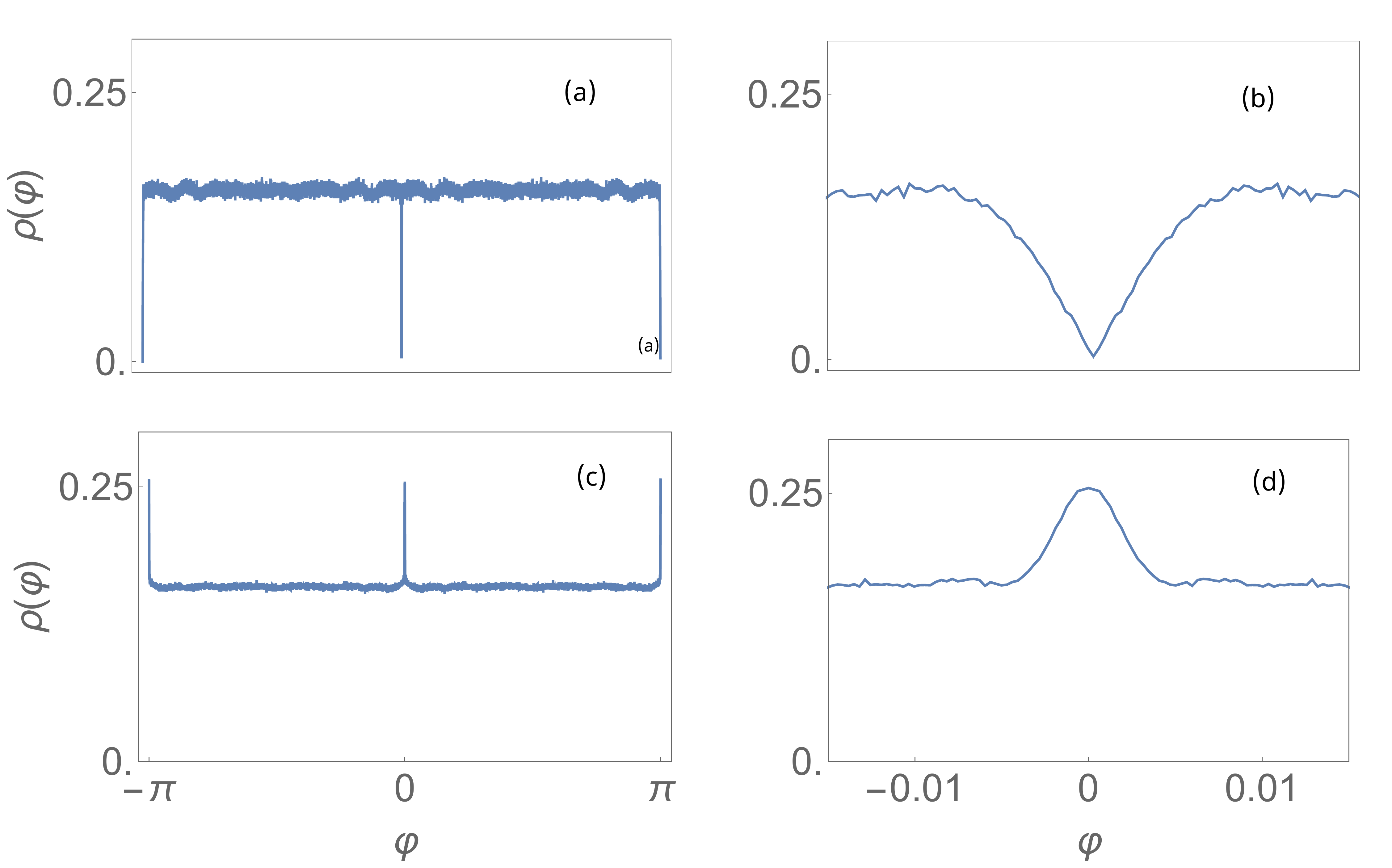}

  \caption{ Disorder averaged eigenphase density for the $\hat{U}_M$, numerical simulation.  Whole support: (a) for $N=9$ and (c) for $N=10$.  Zoom around zero: (b) for $N=9$ and (d) for $N=10$.}
  \label{Fig1213}
 \end{center}
\end{figure}

The existence of the second anti-unitary symmetry $\hat C^N$ means that the expected universality class of the  self-dual  chain in the limit of large $N$ is  not COE but a `'new'' (non-Wigner-Dyson) one. Numerical tests confirm that indeed the  distribution of the eigenphases
of $\hat U_M$ agrees with that of the ensemble T$_+$CQE for $N$ odd (CQE deciphered as the circular quaternion ensemble, with $T_+$ indicating the presence of the time reversal) and
T$_+$CRE (circular real ensemble of orthogonal matrices \footnote{ Not to be mixed with COE of symmetric unitary matrices!}) for $N$ even. Both are characterized by linear eigenphase repulsion, $\beta=1$; the spectacular new element is that the average eigenphase density is no longer constant but has narrow linear minima, $\alpha=1$ (resp. quadratic maxima, $\alpha=0$) at $\phi=0, \pm\pi$ for $N$ odd (resp. even)  \cite{Beena},  see Fig.\ \ref{Fig1213}.
In the vicinity of $0,\;\pm\pi$ the spectrum of  $\hat U_M$ reduces  to that of Hamiltonians of  ``new'' symmetry classes, see Ref.\ \cite{Verbaa,AltZirn,Kieburg}.

\subsection{The dual operator}

The operators $\hat{W}_n$ possess unitary symmetries with respect to cyclic permutations $\hat{P}_C$ and reflection $\hat{P}_R$ with
$\hat{P}_C\left|\sigma_1,\ldots,\sigma_T\right\rangle=\left|\sigma_2,\ldots,\sigma_T,\sigma_1\right\rangle$ and $\hat{P}_R\left|\sigma_1,\ldots,\sigma_T\right\rangle=
\left|\sigma_T,\ldots,\sigma_2,\sigma_1\right\rangle$.  For odd $T$ the  symmetrized basis
\begin{eqnarray}\label{eta}
|\eta^{(k)}_\alpha\rangle&=&A\sum_{t=0}^{T-1}\hat{P}_C^t{\rm e}^{\frac{2\pi ikt}{T}}|\vec{\sigma}^T_\alpha\rangle,\hspace*{2mm}k=1,\ldots,T-1\nonumber\\
|\eta^{\left({0\pm}\right)}_\alpha\rangle&=&A'(1\pm\hat{P}_R)\sum_{t=0}^{T-1}\hat{P}_C^t
|\vec{\sigma}^T_\alpha\rangle,\hspace*{2mm}k=0
\end{eqnarray}
% with normalizing constants $A$, $A'$
 allows to decompose $\hat{W}_n$ into $T+1$ irreducible blocks  $\hat{W}_n^{(k)}$; if $T$ is even the block with $k=T/2$ falls
 into $\pm$-subblocks and the overall number of irreducible blocks is $T+2$. Note that
 \begin{eqnarray*}
 \hat{P}_C|\eta^{(k)}_\alpha\rangle&=& {\rm e}^{\frac{2\pi ik}{T}}|\eta^{(k)}_\alpha\rangle,\\
 \hat{P}_C\left(\hat{P}_R\,|\eta^{(k)}_\alpha\rangle \right)&=& {\rm e}^{\frac{2\pi i(T-k)}{T}}\left(\hat{P}_R\, |\eta^{(k)}_\alpha\rangle\right)
 \end{eqnarray*}
such that $\hat{P}_R$ changes $k\to T-k$. It follows that with the proper choice of the phases of the symmetrized basis functions we have   $\hat{W}_n^{(k)}=\hat{W}_n^{(T-k)}$.

If $T$ is prime all $T$ states obtained from $|\vec{\sigma}^T_\alpha\rangle$ by consecutive cyclic shifts are different, unless  $|\vec{\sigma}^T_\alpha\rangle$ is either $|1,1,\ldots,1\rangle$ or $|-1,-1,\ldots,-1\rangle$. It follows that the dimensions  $M_k$ of the symmetry blocks $\hat{W}_n^{(k)}$ then obey,
\begin{eqnarray}
M_k=\frac{2^T-2}{T},\quad 0<k<T,\nonumber\\
M_{0+}+M_{0-}=2+\frac{2^T-2}{T}.
\end{eqnarray}
The dimensions of the $0\pm$-subspaces  are given in the Appendix.

Passing to anti-unitary symmetries, we note that in the dual approach time and the particle number are effectively interchanged; since the set of disordered fields  $h_n$ is not assumed symmetric, there is no operator analogous to the time reversal. On the other hand, there exists the identity analogous to (\ref{symmrelU}) connecting the dual operators $W'_n$ for $J=\pi/4+\Delta J$ and $W_n$ for  $J=\pi/4-\Delta J$,
\begin{eqnarray}
 \hat{W}_n'=(-i)^T \hat{\Sigma}_y^T \hat{W}_n^*\hat{\Sigma}_y^T,\quad \hat{\Sigma}_y^{T}=\otimes_{t=1}^{T}\hat{\sigma}_t^y\label{symmrelW};
\end{eqnarray}
we will prove it here by considering the matrix elements in Eq.\ (\ref{dualop}). For the left hand side of the relation we get
\begin{equation}\label{matr1}
 \left\langle\vec{\sigma}^{T}{'}|\hat{W}_I{'}|\vec{\sigma}^{T}\right\rangle\left\langle\vec{\sigma}^{T}|\hat{W}_{n,b}|\vec{\sigma}^{T}\right\rangle\, .
\end{equation}
Similar to  (\ref{sigmamatrelt}), only the matrix elements $\left\langle\vec{\sigma}^{T}|\Sigma_y^T|-\vec{\sigma}^{T}\right\rangle$ can be non-zero such that the right hand side can be written,
\begin{eqnarray}\label{matr3}
(-i)^T \eta(\vec{\sigma}^{T}{'})\eta( \vec{\sigma}^{T})\\
\times\left\langle -\vec{\sigma}^{T}{'}|\hat{W}_I^{*}|-\vec{\sigma}^{T}\right\rangle\left\langle-\vec{\sigma}^{T}|\hat{W}_{n,b}^*|-\vec{\sigma}^{T}\right\rangle \nonumber
\end{eqnarray}
where $\eta( \vec{\sigma}^{T})$ is the number of spins-up in $\vec{\sigma}^T$. The matrix elements of $\hat{W}_{n,b}^*$ in Eq.\ (\ref{matr3}) and of $\hat{W}_{n,b}$ in Eq.\ (\ref{matr1}) are equal (the minus sign of $\vec{\sigma}^{T}$ is compensated by complex conjugation in the $h_n$-dependent factor and doesn't influence the real $R$-dependent factor determined by the number of the domain walls in   $\vec{\sigma}^{T}$).
Denoting by $\mu$ the number of positions where the entries in $|\vec{\sigma}^{T}{'}\rangle$ and $|\vec{\sigma}^{T}\rangle$ are different, we get for the matrix element of $\hat{W}_I^*$ in Eq.\ (\ref{matr3})
${\exp}\left(i(\pi/4-\Delta J)(T-2\mu)\right)$ and for the matrix elements of $\hat{W}_I{'}$ in Eq.\ (\ref{matr1}) ${\exp}\left(-i(\pi/4+\Delta J)(T-2\mu)\right)$.
It can be seen that the $\mu$-dependent factors cancel with $\eta( \vec{\sigma}^{T}) \eta(\vec{\sigma}^{T}{'})$  and the relation (\ref{symmrelW}) follows.

In the self-dual case, defining $\hat W_{n,M}\equiv \exp\left(iT\pi/4\right)\hat{W_{n}}$ we have the ``charge conjugation'' property, $\quad \left[\hat W_{n,M},\hat C^T\right]=0$ where
\begin{eqnarray}
\label{charconjW}
\hat C^{T}=\hat{\Sigma}_y^{T}\hat K,\qquad \hat C^2=(-1)^{T}.
\end{eqnarray}
Same reasoning as for the evolution operator shows that  $\hat W_{n,M}$  are symplectic if $T$ is odd and orthogonal if $T$ is even.

\begin{figure}[b]
 \begin{center}
 \hspace{-0.6cm}
  \includegraphics[width=9cm]{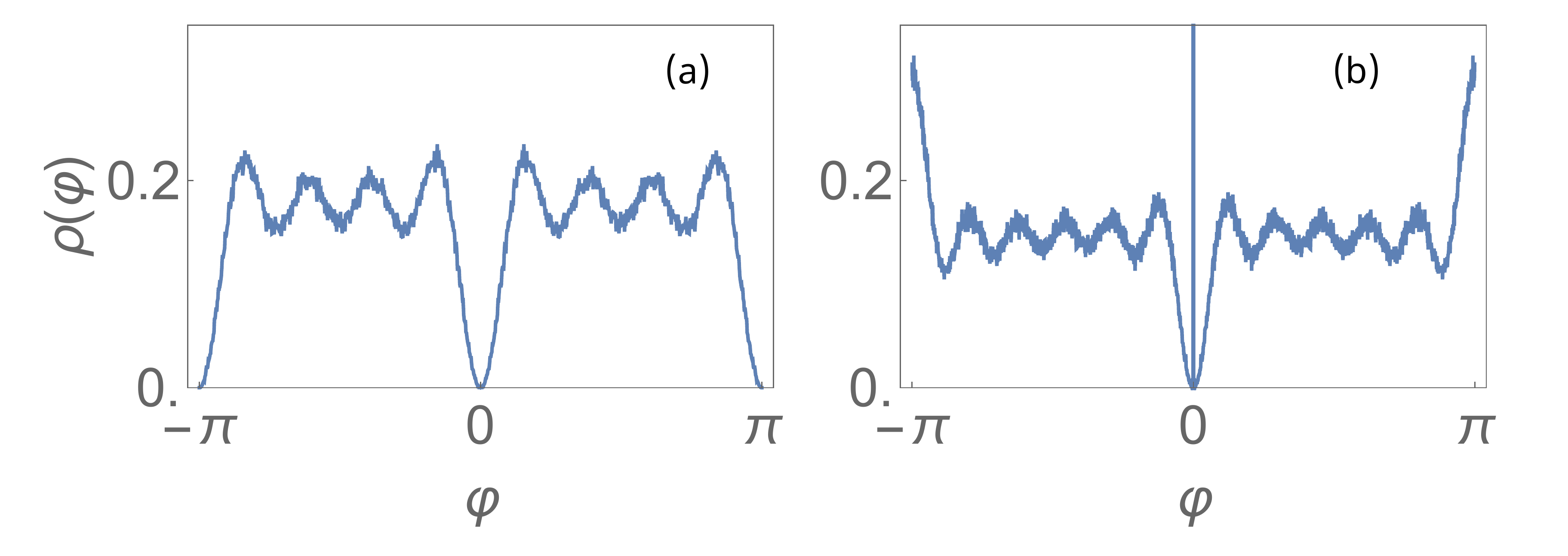}
\caption{Disorder averaged eigenphase density $\rho(\varphi)$ of the  block $0+$ of $\hat W_M$ for $N=13$, $T=5$ (a) and $T=6$ (b). The delta like peak at zero at the  plot (b)
indicates the zero mode. This corresponds to Eq.\ (\ref{denss}) for $n=9$ and $n=12$, respectively, after proper normalization.}
  \label{Fig1011}
 \end{center}
\end{figure}

In sufficiently long chains the spectra of the global dual operators (products of the local random $\hat W_n$) tend to universality because of the Furstenberg-like mechanism \cite{Haake}. To formulate their properties we must get rid of the unitary symmetries and consider the irreducible blocks  $\hat W^{(k)}_{n,M},\quad  \hat W^{(k)}_M$.
In Fig.\ \ref{Fig1011}
we show the numerically computed eigenphase densities $\rho(\varphi)$
%obtained after desymmetrization for
of the block $0+$ of $\hat{W}_M$ for different $T$. For sufficiently large $N$ we find
perfect agreement with the predictions by the following ``new'' ensembles of RMT:
the circular quaternion ensemble (CQE) for $T$ odd and the circular real ensemble (CRE) for $T$ even. Both are characterized by quadratic level repulsion $\beta=2$, a quadratic behaviour
around zero and $\pm\pi$ and the density \cite{Wiki,Girko}
\begin{equation}\label{denss}
\rho(\varphi)=1\pm\frac{\sin n\varphi}{n\sin\varphi}.
\end{equation}
For odd $T$  only the minus sign is realized with odd $n=\text{dim}\;\hat W^{(k)}+1$. For even $T$  both signs  occur and
$n=\text{dim}\;\hat W^{(k)}-1$ can be even or odd such that $\rho(\varphi)$ can have a minimum or a maximum in the vicinity of $0$ and $\pm\pi$; at the points of minima disorder-protected eigenphases  similar to the Majorana zero modes \cite{Beena} appear in the spectrum, see Fig.\ \ref{Fig1011}(b).

{\section{Spectral Form Factor}}  The quantity of interest is the spectral form factor
\begin{equation}
 K_{N}\left(  T\right)=\left\langle\left|\textrm{Tr\,}\hat{U}^T\right|^2\right\rangle=\left\langle\left|\operatorname{Tr}\,\prod_{n=1}^N\hat{W}_{n}\right|^2\right\rangle,
\end{equation}
where $\left\langle\ldots\right\rangle$ denotes the disorder average over $h_n$. As shown in  \cite{Prosen} , if the  $h_n$ are assumed independent and Gaussian-distributed, averaging can be carried out analytically. Consider  the squared dual Hilbert space  with dimension $2^{2T}$ and the product basis
$\left|\vec{\sigma}\vec{\sigma}'\right\rangle=\left|\vec{\sigma}^T \right\rangle\otimes\left|\vec{\sigma}^T{'}\right\rangle$;
operators in that space will be denoted by calligraphic letters. 
The form factor can then be written,
\begin{equation}\label{fordis0}
K_N(T)={\rm Tr}\,\hat{\mathcal{A}}_\xi^N, \hspace*{3mm} \hat{\mathcal{A}}_\xi\equiv(\widehat{\overline{W}}\otimes\widehat{\overline{W}}^*)\hat{\mathcal{O}}_\xi
\end{equation}
with $\widehat{\overline{W}}$ obtained from $\hat{W}$ by the replacing $h_n$ by its average.
The nonzero matrix elements of $\hat{\mathcal{O}}_\xi$ are given by
\begin{equation}
\langle \vec{\sigma}\vec{\sigma}^{\prime}| \hat{\mathcal{O}}_{\xi
}| \vec{\sigma}\vec{\sigma}^{\prime}\rangle =\exp\left[
-\frac{\xi^{2}}{2}\left(  \sum_{t=1}^{T}\sigma_{t}-\sum_{t=1}^{T}%
\sigma_{t}^{^{\prime}}\right)  ^{2}\right]
\end{equation}
where $\xi$ stands for the standard deviation of $h_n$.
Here we do not consider the crossover between regularity and disorder and  study only the limit of  large $\xi$ \cite{Valid}. Then $\hat{\mathcal{O}}_\xi$ can be replaced by a projector $\hat{\mathcal{P}}$ with unit matrix elements
for $\sum_{t=1}^T\sigma_t=\sum_{t=1}^T\sigma_t'$ and zero otherwise;
the form factor reduces to,
\begin{equation}\label{fordis}
K_N(T)={\rm Tr}\,\hat{\mathcal{A}}^N,
\hspace*{3mm} \hat{\mathcal{A}}\equiv \hat{\mathcal{P}}(\widehat{\overline{W}}\otimes\widehat{\overline{W}}^*)\hat{\mathcal{P}}.
\end{equation}
Using the block diagonal structure of  $\widehat{\overline{W}}=\oplus_{k}\widehat{\overline{W}}^{(k)}$ and $\hat{\mathcal{A}}=\oplus_{kk'}\hat{{\mathcal{A}}}^{(kk')}$ in the basis $|\eta^{(k)}_\alpha\rangle\otimes|\eta^{(k')}_\alpha\rangle^*$ and denoting
the eigenvalues
of $\hat{\mathcal{A}}^{(kk')}$  by $\lambda_j^{(kk')}$ we have,
\begin{equation}\label{sumlam}
K_N(T)=\sum_{kk'}\sum_j\left[\lambda_j^{(kk')}\right]^N.
\end{equation}

The basis truncation by $\hat{\mathcal{P}}$  shifts the vast majority of  eigenvalues  of  the unitary operator $\widehat{\overline{W}}\otimes\widehat{\overline{W}}^* $ inside the unit  circle such that their contribution in the thermodynamic limit $N\to \infty$ dies out. Some of the $\lambda_j^{(kk')}=\pm 1$ are unchanged by the basis truncation, and it is these eigenvalues which define the form factor of long chains.   It is elementary to show that as a consequence of unitarity of  $\widehat{\overline{W}}^{(k)}$, 
 each diagonal block  $\hat{{\mathcal{A}}}^{(kk)}$ does have an eigenvalue unity associated with the  eigenvector   $\sum_{\alpha}\vert
\eta_{\alpha}^{\left(  k\right)  }\rangle\otimes \vert \eta_{\alpha
}^{\left(  k\right)  }\rangle ^{\ast}$;  we shall refer to it as the trivial eigenvector. In view of $\hat{W}^{\left(  k\right)  }=\hat{W}^{\left(  T-k\right)  }$  the eigenvalue 1 is also present in the spectra of $\hat{{\mathcal{A}}}^{(k,T-k)},\quad 0<k<T$.
Counting the relevant symmetry blocks of $\mathcal{A}$ we obtain that  in the self-dual case and $T>5$   it has $2T$  eigenvalues unity associated with the trivial eigenvectors \footnote{For $t\le 5$ the representation $0-$ is absent such that the number of trivial eigevectors is one less}.  In the limit $N\to \infty$ they give rise to the form factor $K=2T$ in line with the
RMT predictions for COE. This result is based on unitarity and symmetry of the dual problem, and thus applies also to other self-dual systems. The trivial $2T$ eigenvalues  unity are the only ones if $T$ is odd while for even $T$ there is one more  such that the thermodynamic limit of the form factor  is $2T+1$; for several small $T$ other number-theoretical peculiarities of $K(T)$ are observed \cite{Prosen}.

\begin{figure}\begin{center}
 \includegraphics[width=9cm,angle=0]{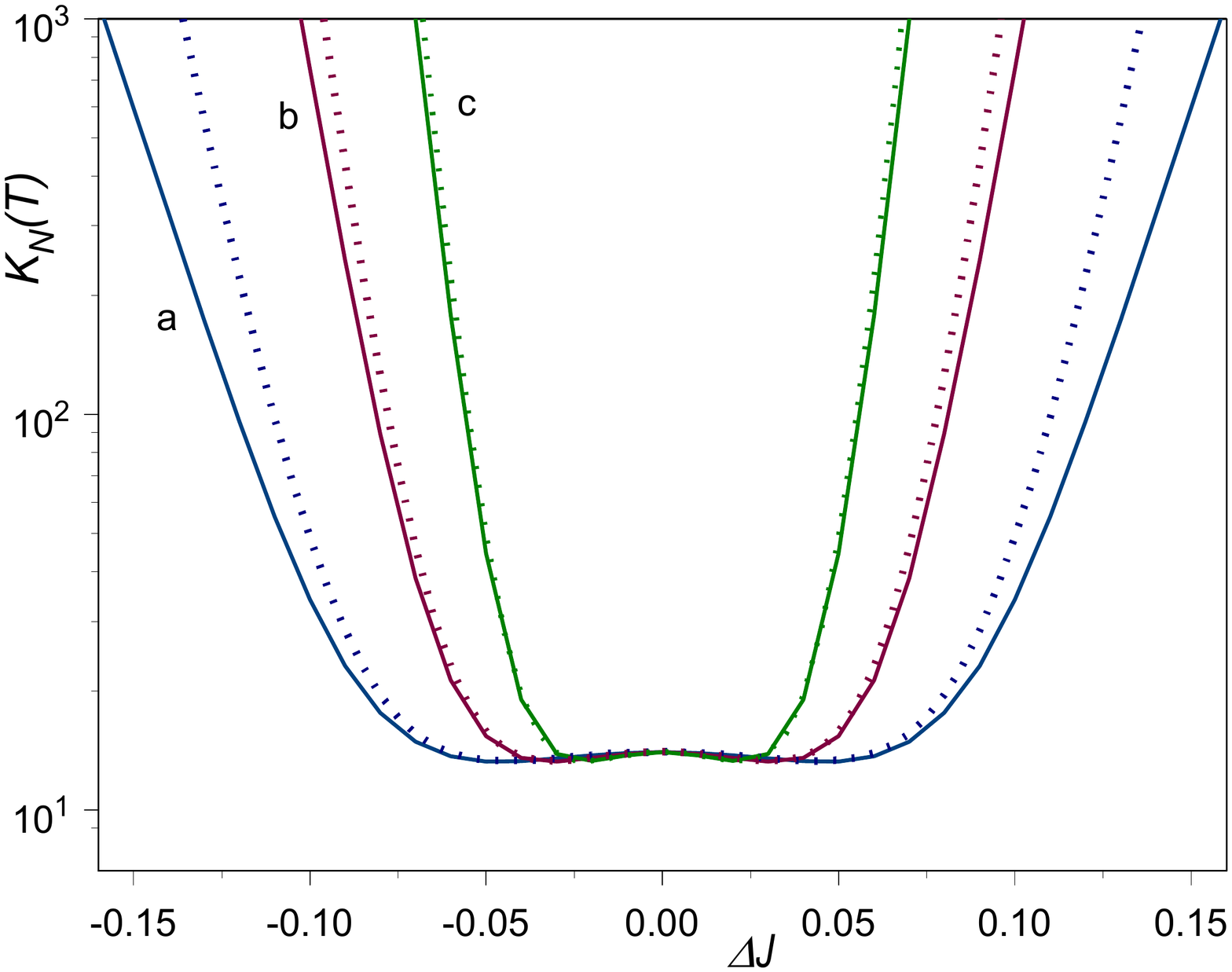}\end{center}
  \caption{Form factor for fixed $T=7$ versus $\Delta J$ for a) $N=40$, b) $N=80$, c) $N=160$.
  Full lines are exact numerics, dotted the approximations (\ref{ffc}).}
  \label{Fig 0}
 \end{figure}

\begin{figure}\begin{center}
 \includegraphics[width=7cm]{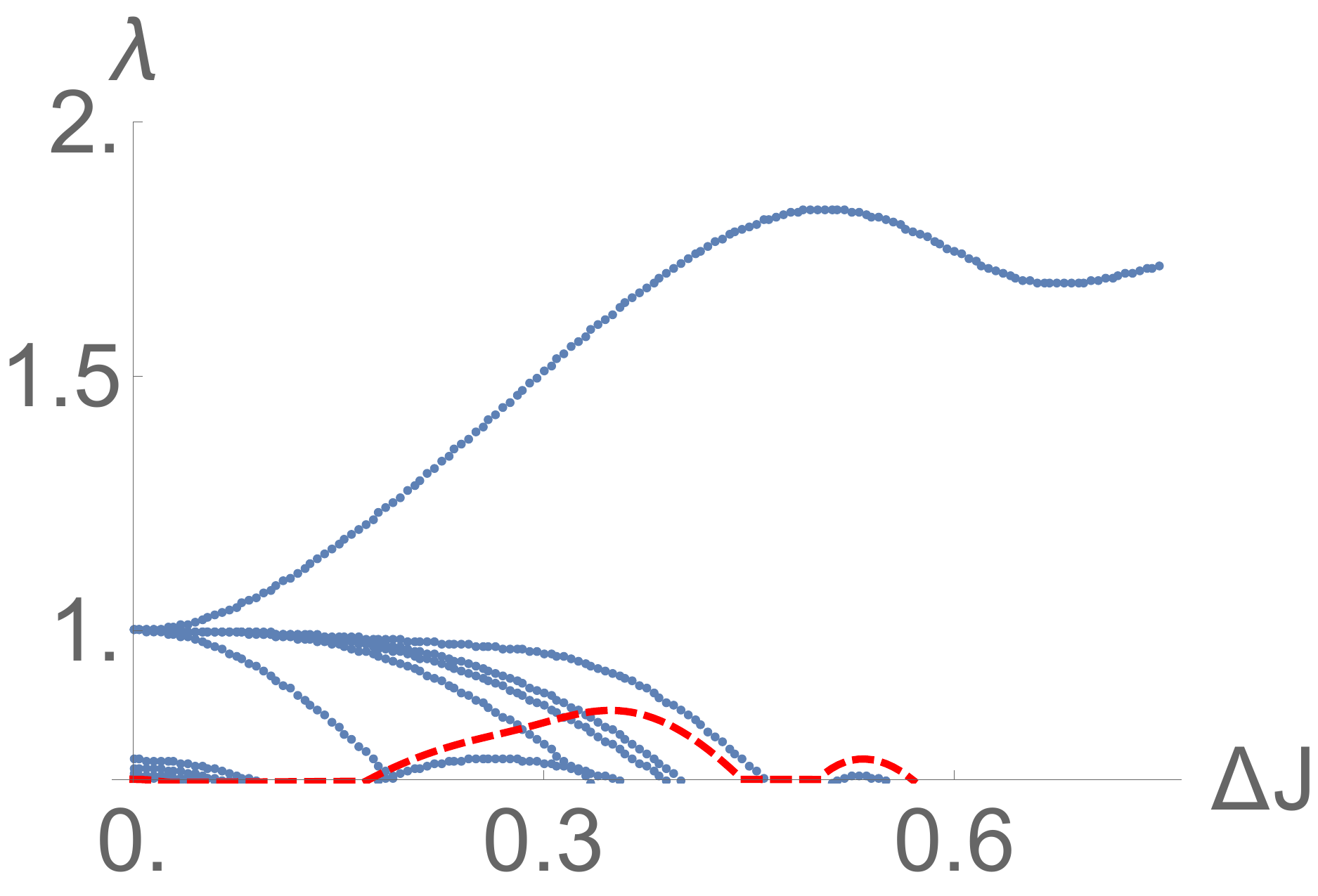}
 \caption{Largest eigenvalues of $\hat{\mathcal A}$ as functions of $\Delta J$, case $T=9$. Dashed (red) lines indicate the absolute value of  complex eigenvalues. Only $\lambda_{0+}$ grows above 1.}
 \label{eivalues}\end{center}
\end{figure}

\begin{figure}
 \includegraphics[width=8cm]{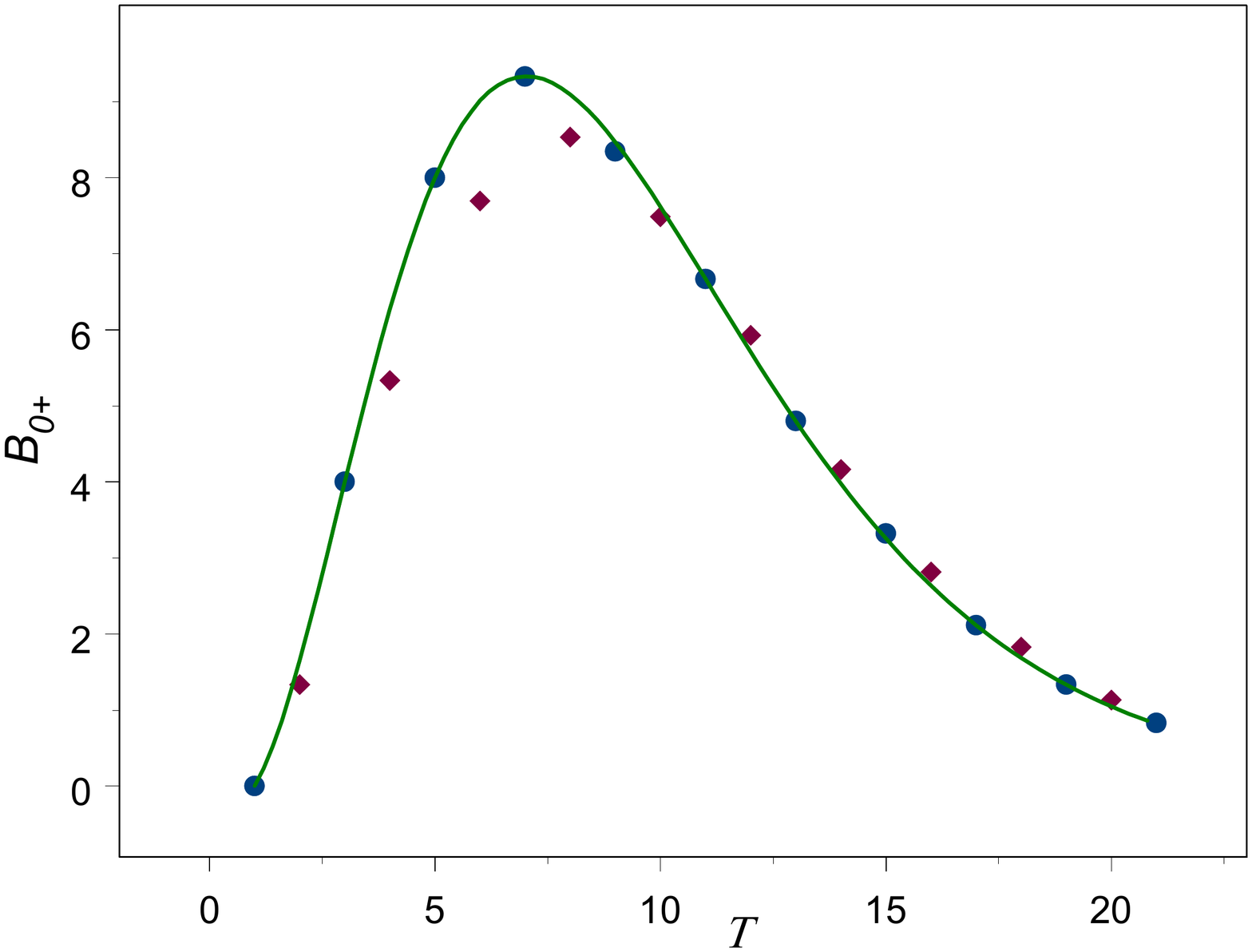}
 \caption{Coefficient $B_{0+}$ versus $T$. The full line is Eq.\ (\ref{bk}),
 the circles (squares) show the exact numerical result for odd (even) $T$.}\label{fig1}
\end{figure}

{\section{Deviation from Self-Duality. Dependence of the Spectral Form Factor on $\Delta J$ and $T$}}  The numerically calculated dependence of $K_N(T)$ on $\Delta J$ is shown for different $N$ in Fig.\ \ref{Fig 0}.
%This behavior was described in the RMT context in Ref.\ characterizing topologically protected subgap states in superconductors.
The relation   (\ref{symmrelU}) implies symmetry of the plot around $J=\pi/4$. When we move away from $J=\pi/4$ the form factor first slightly decreases, then forms a plateau and
finally increases exponentially. In the limit $N\to\infty$ the plateau shrinks while the increase becomes ever sharper
reminiscent of a phase transition. According to Eq.\ (\ref{sumlam}) the dominant contributions to the form factor of long chains result only from the largest eigenvalues that smoothly transform into the unit eigenvalues in the self-dual case; see Fig.~\ref{eivalues}.
 In the interval of the exponential growth of $K_N(T)$ the situation further simplifies, as
the only significant contribution results from the largest eigenvalue of the block $\hat{\mathcal{A}}^{(0+,0+)}$. Due to Eq.\ (\ref{symmrelW}), $\lambda_k=\text{max}_j\,\lambda_j^{(kk)}$ is an even function of $\Delta J$,
thus
\begin{align}
\lambda_{k}  & =1+B_{k}\Delta J^{2}+O\left(\Delta J^{4}\right).
\end{align}
We obtained  the coefficients $B_k$ for prime $T$,
\begin{eqnarray}\label{bk}
B_k&=& -\frac{2T(T-1)}{2^{T-1}-1},\hspace*{3mm}B_{0\pm}=\pm\frac{2T(T-1)}{2^{(T-1)/2}\pm1}
%B_{0u}&=&-\frac{2T(T-1)}{2^{(T-1)/2}-1}\hspace*{3mm}{\text{for}}\hspace*{3mm}T>5\hspace*{3mm}\hspace*{3mm}0\hspace*{3mm}\text{otherwise}.\nonumber
\end{eqnarray}
with $B_{0-}=0$ for $T\leq5$. Derivation of these formulae is given  in the Appendix. 
We  first examine how good our analytical expressions for $B_k$ are for other values of $T$.
We concentrate on the coefficient $\lambda_{0+}$ that we find dominant in the
long chain limit $N\to\infty$.
Therefore we plot in Fig.\ \ref{fig1}   the expression in Eq.\ (\ref{bk}) for the
coefficient $B_{0+}$ as full line and  the results from full numerical calculations for all integer $T$ as dots.
We observe that the expression (\ref{bk}) provides a good interpolation also for nonprime $T$-values as
well, especially for the odd ones.

%Although these expressions are exact for prime $T$ they provide a good interpolation formula also for other $T$, see Fig.\ 1 of Ref.\ \cite{suppl}.
%There is excellent agreement for $T$ odd, the agreement for even $T$ is rather good for larger $T$.
In the limit $N\to\infty$, $\Delta J\to0,$ $N\Delta J
^{2}\equiv x=\mathrm{const.}$, the relation $\lambda_{k}^{N}\approx\exp\left(  N\Delta J^{2}B_{k}\right)$ becomes exact yielding the spectral form factor
\begin{equation}\label{ffc}
 K_N(T)={\rm e}^{xB_{0+}}+{\rm e}^{xB_{0-}}+2(T-1){\rm e}^{xB_{k}}.
\end{equation}
We find a good agreement that improves with increasing $N$ between this expression and the exact numerical result as shown in Fig.\ \ref{Fig 0}.
%\begin{figure}
%[ptb]
%\begin{center}
%\includegraphics[height=7cm]
%{GS1.pdf}%
%\caption{ Exact form factor (full line) and its \textquotedblleft approximation by
%exponentials\textquotedblright (dotted line) for $T=11$ and $N=80$.}%
%\label{Fig.8}%
%\end{center}
%\end{figure}
The slight decrease of $K_N(T)$ in the vicinity of $\Delta J=0$ can be traced back to
the fact that the sum of all $B_k$, $B_{0\pm}$ is slightly smaller than zero, the exponential growth is determined by $B_{0+}$.
Expression (\ref{ffc}) shows the correct RMT behavior $K_{N}\left(  T\right)  =2T$
for fixed $x$ and large $T$ when all $B_{k}$ become exponentially small.

{\section{Issue of Many-Body Localization}}  
\subsection{Small $\Delta J$: Perturbation Theory Domain}
\begin{figure}
[ptb]
\begin{center}
\includegraphics[height=7cm]
{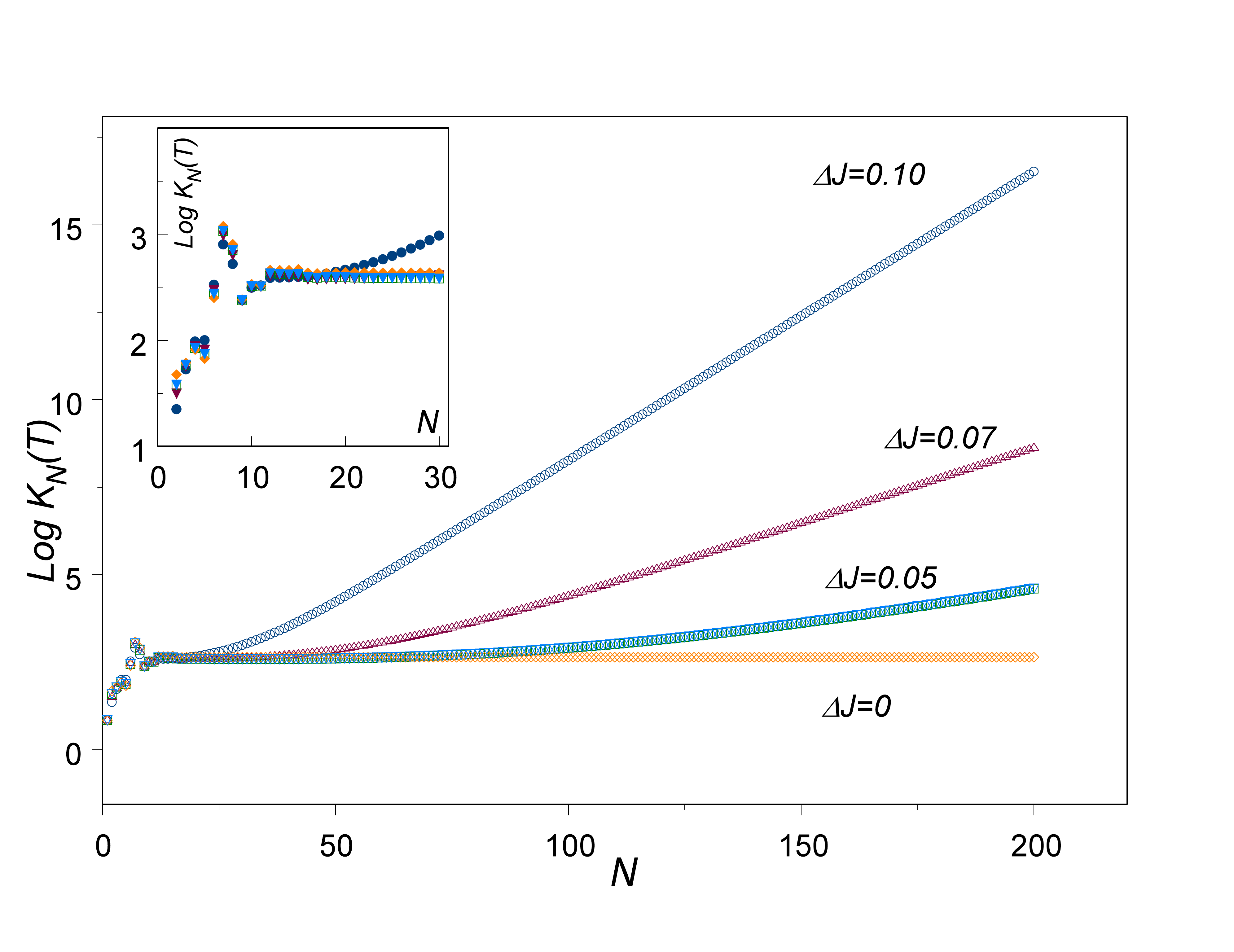}%
\caption{Logarithm of the form factor for $T=7$ time steps versus the chain length $N$.
The erratic small $N$ behavior is shown in the inset.}%
\label{Fig. 3}%
\end{center}
\end{figure}

We plot in Fig.\ \ref{Fig. 3} $K_N(T)$  as function of $N$.
After an erratic behavior for small $N$ and a  RMT transition region we find an exponential growth with $N$. This hints at a localization effect. To clarify this, we
consider a toy model of a set of $\mu$ chaotic non-interacting non-identical
quantum systems each belonging to a RMT universality class, e.g.\ the COE with the form factor $K_{\rm{COE}}(T)$. The energy
spectrum of the whole set will be a direct sum of the spectra of separate
systems and the form factor will be the
$\mu$-th power of $K_{\rm{COE}}(T)$.
\iffalse
%
\begin{align}\label{loc}
K_N\left(  T\right)   & =\left\langle
{\displaystyle\sum\limits_{k_{1}l_{1}}}
{\rm e}^{i\left(  \varphi_{k_{1}}-\varphi_{l_{1}}\right)  T}\right\rangle \ldots\left\langle
\sum_{k_{\mu}l_{\mu}}{\rm e}^{i\left(  \varphi_{k_{\mu}}-\varphi_{l_{\mu}}\right)
T}\right\rangle \nonumber\\
& =\left(  K_{\rm{COE}}\left(  T\right)  \right)  ^{\mu}.
\end{align}
\fi

Now consider a disordered chain of $N$ spins suspected to undergo
localization, with localization length of some $N_{c}$. For
$N<N_{c}$ the chain length is too small for the localization to occur and the form
factor observed is $K_{\rm{COE}}(T)$. For larger $N$ the eigenphase spectrum will be a direct sum of $N/N_{c}=\mu$ spectra with the form
factor $\left(  K_{\rm{COE}}\left(  T\right)  \right)  ^{\mu}$.
%\begin{equation}
%K_N\left(  T\right)  =  K_{\rm{COE}}\left(  T\right)    ^{\mu},\quad
%N>N_{c}.
%\end{equation}
The plot of $\ln K_N\left(  T\right)  $ as function of the
chain length $N$ would thus consists of a horizontal stretch at $N<N_{c}$ and a tilted
line with inclination $\tan\phi=\ln K_{\rm{COE}}\left(  T\right)  /N_{c}$ for
$N>N_{c}$. %The localization length would be the abscissa of the crossing of the
%large-$N$ asymptotics of $\ln K_N\left(  T\right)$ with the line $y=\ln
%K_{\rm{COE}}\left(T\right)$.

\begin{figure}\begin{center}
 \includegraphics[width=8cm]{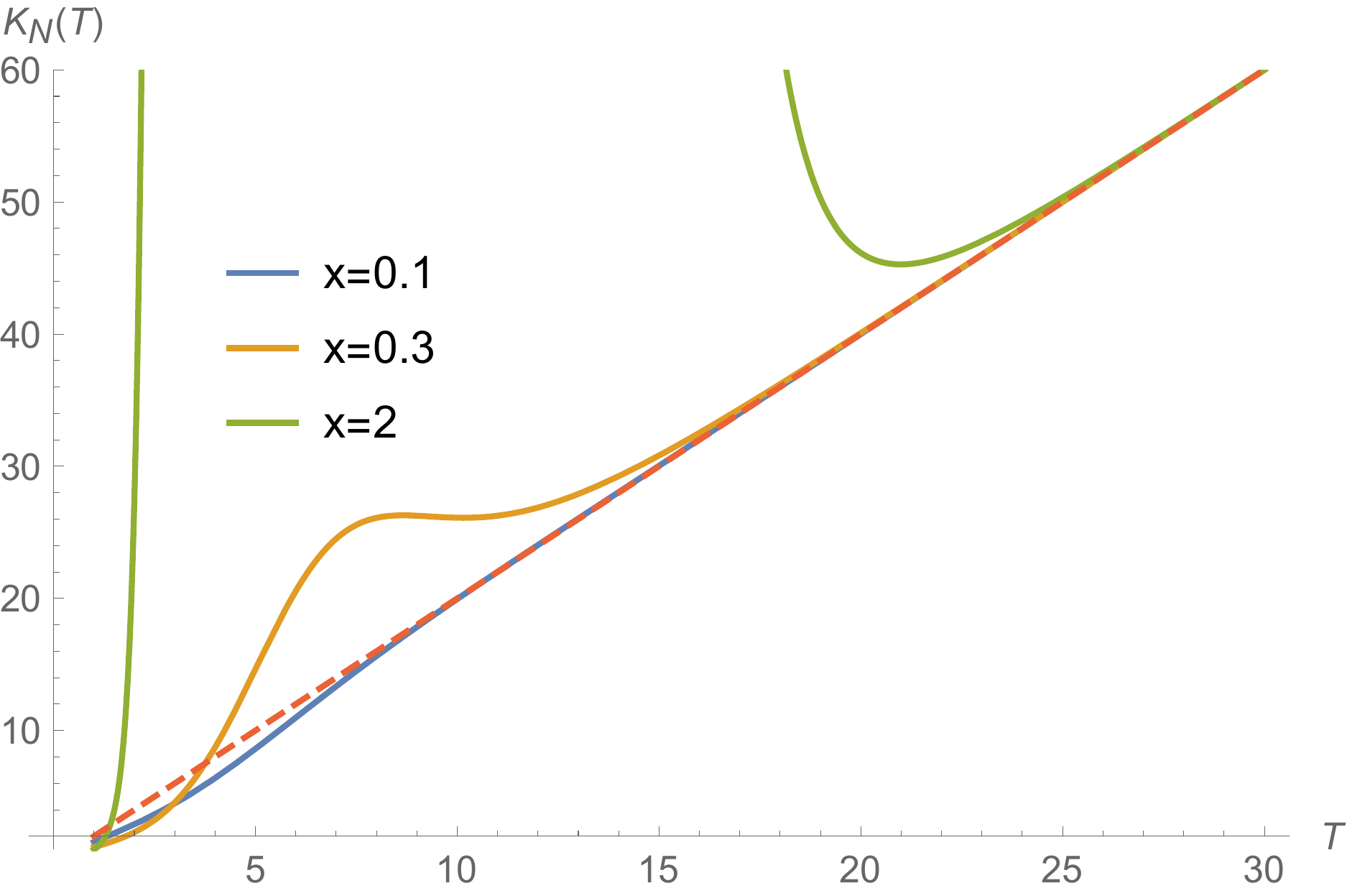}
\caption{The perturbation theory form factor compared with the RMT prediction (dashed).
The deviation tends to zero at $T>T_{Th}$ (see Eqs.\ (\ref{ffc},\ref{neweq})) approximately equal to $10,17$ and $25$,
respectively. The upper line is obtained for $x=2$, the middle one for $x=0.3$ and the lower one for $x=0.1$.}\label{newfig}\end{center}
 \end{figure}

Returning to the system under discussion, we recall that in the limit of long
chains the form factor reduces to the $N-$th power of the largest eigenvalue
$\lambda_{0+}$ of $\hat{\mathcal{A}}$. Consequently the time dependent localization
length is found as $N_{c}(T)={\ln K_{\rm{COE}}}/{\ln\lambda_{0+}}\approx\ln 2T/(B_{0+}\Delta J^2)$.
For a fixed $x=N\,\Delta J^{2}$ and time growing the perturbation theory
form factor (\ref{ffc}) first undergoes strong non-monotonic changes and then
stabilizes at the RMT value $2T$ (Fig.\ \ref{newfig}).
The stabilization threshold can be
called the Thouless time of the problem $T_{Th}$; its dependence on $x$
follows from (\ref{ffc}) if we demand that the difference of $K_N\left( T\right) $ from
$2T$ is much smaller than $2T$. For $x\gtrsim 0.3$ all summands in the
stabilization condition but ${\rm e}^{xB_{0+}\left( T\right) }$ can be neglected
and we get the Thouless time from the boundary
\begin{equation}\label{neweq}
\frac{\log 2T}{B_{0+}\left( T\right) }\gtrsim N\,\Delta J^{2}.
\end{equation}
\subsection{Large $\Delta J$}
\begin{figure}\begin{center}
\includegraphics[width=9cm]{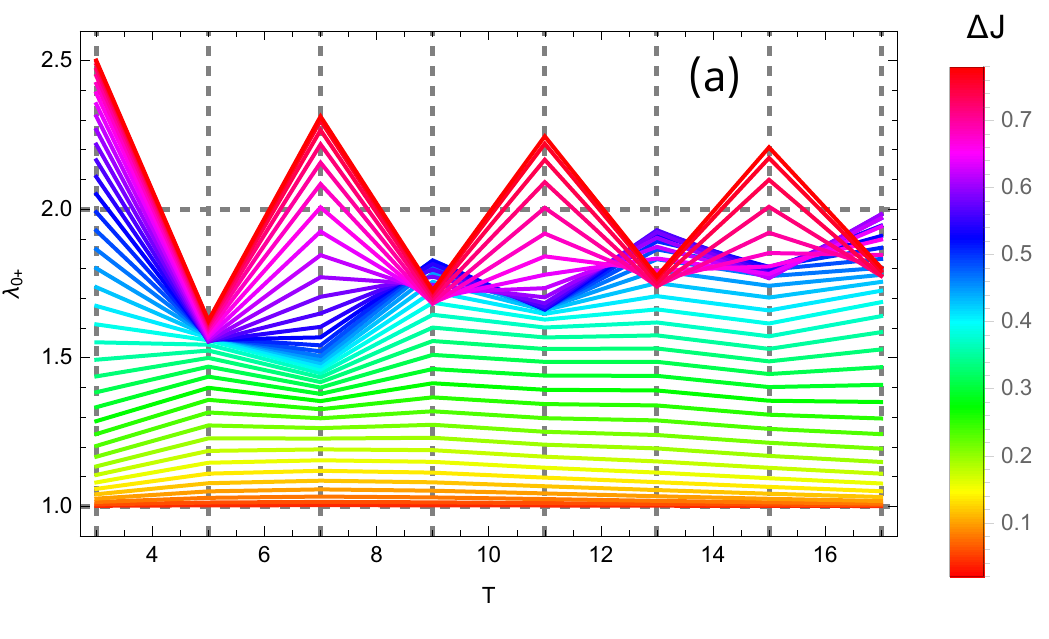}
\includegraphics[width=9cm]{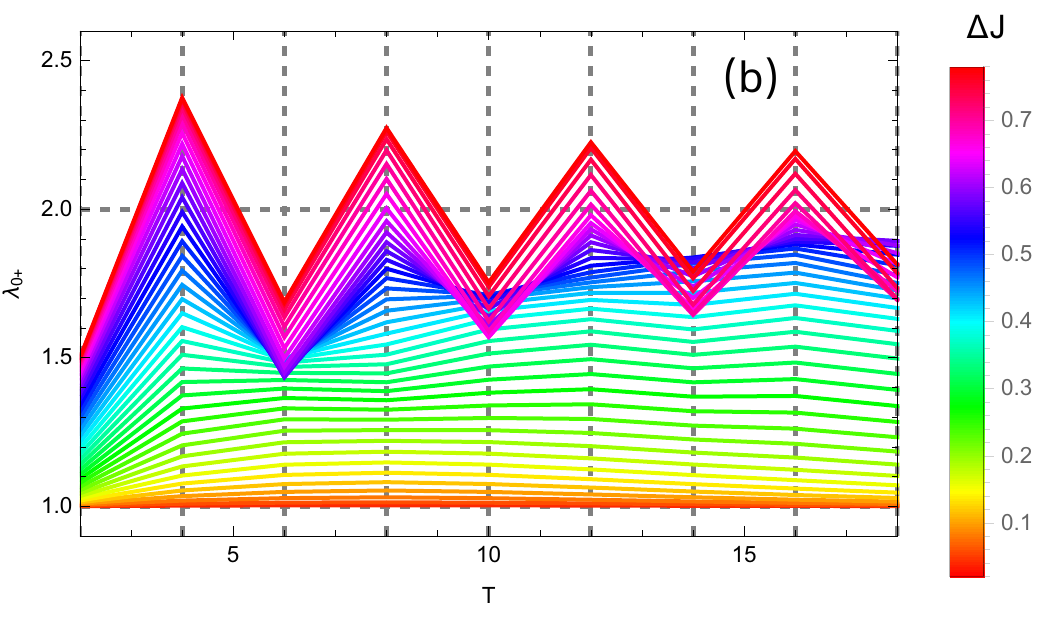}
\caption{Coefficient $\lambda_{0+}$ versus $T$ for $T$ odd (a) and $T$ even (b) with color coded values of $\Delta J$.}
\label{fig}\end{center}
\end{figure}

Whereas the latter relation is restricted to the regime of small $\Delta J$,
our dual operator approach allows to decide upon localization or ergodic behavior for general $\Delta J$. We
show $\lambda_{0+}$ in Fig.\ \ref{fig} in dependence of odd and even $T$
separately.
for the full range of $\Delta J$ where the
system undergoes a transition from the behavior around $\Delta J = 0$, where the eigenvalue
$\lambda_{0+}$ is almost
flat, to the integrable case  $\Delta J=\pi/4$, or  $J=0,\pi$, marked
by strong oscillations. We remind that in view of the symmetry (\ref{symmrelU}) the limit $\Delta J=\pi/4$ is equivalent to $J=0$, i. e., the chain of non-interacting spins. In-between we find a qualitative
change in the behavior of the eigenvalue as its decay  is first drastically diminished,
then oscillations occur with the amplitude decreasing with time. Remarkably, at any second
data point, $\lambda_{0+}$ grows with $\Delta J$ up to $\Delta J\approx 0.6$ and then decreases
leading to counteroscillating curves for $\Delta J$ smaller and larger than $0.6$.

We thus expect that for large times and
$\Delta J$ above a certain threshold, the decay of oscillations
stop
% for all $T$
such that $N_c(T)$ tends to a limit
smaller than $N$ indicating localization. Although this dual-operator study allows
to consider arbitrary large $N$, it is restricted
to finite $T$-values. For a complementary analysis that permits to study arbitrary $T$
for finite $N$-values
we consider first the average spacing ratio $\langle r\rangle =
\langle \text{min} (\varphi_{n+1}-\varphi_n,\varphi_{n}-\varphi_{n-1})/
\text{max} (\varphi_{n+1}-\varphi_n,\varphi_{n}-\varphi_{n-1})$ where $\varphi_{n-1}$, $\varphi_{n}$,
$\varphi_{n+1}$ are adjacent eigenphases \cite{Hose}.
\begin{figure}\begin{center}
 \includegraphics[width=8cm]{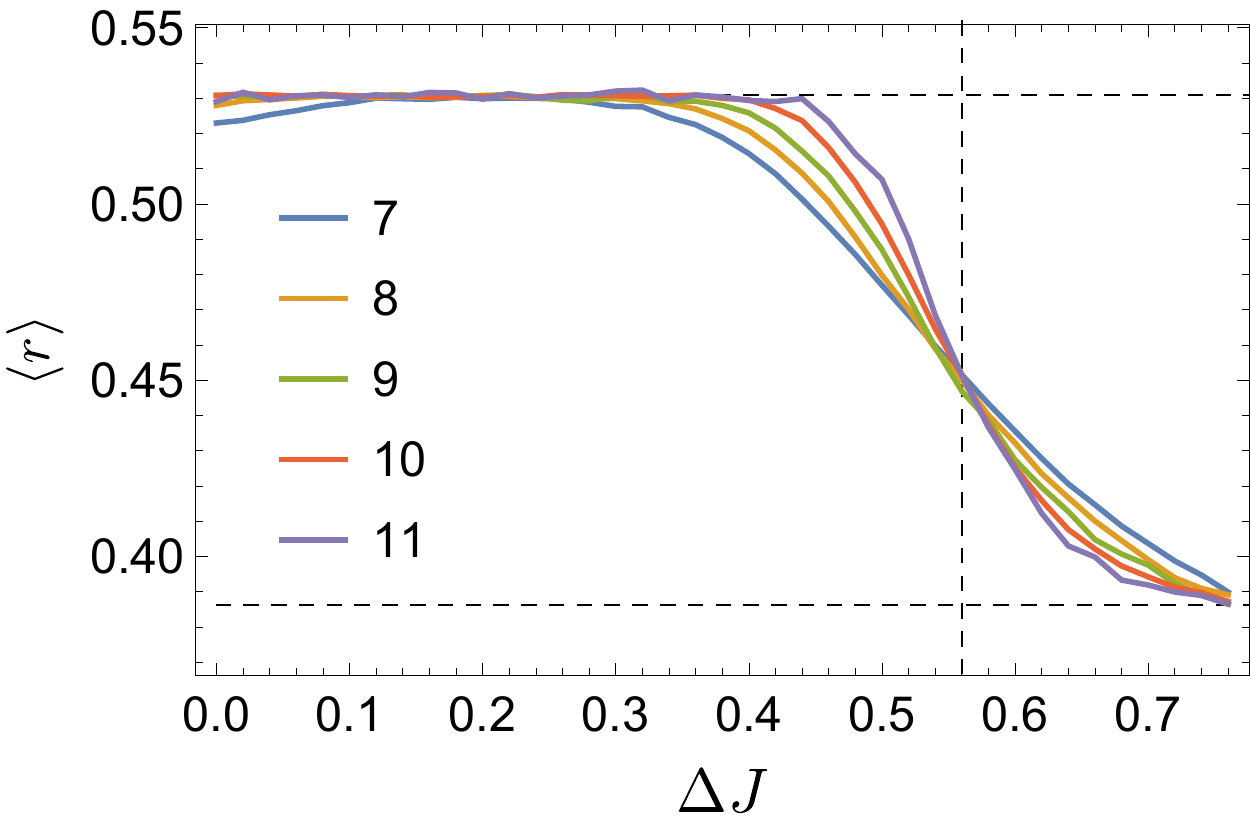}
 \caption{Ratio of eigenphase spacings $\left\langle r\right\rangle$ over $\Delta J$ for different particle numbers $N$. Upper (lower) horizontal lines show the expected behavior for ergodic (localized) behavior.
 The vertical line marks the transition. The lines show $\langle r\rangle$ ordered, left of the transition point, from $N=7$ (bottom) to $N=11$ (top).}
 \label{rstat}\end{center}
\end{figure}
It is shown in Fig.\ \ref{rstat} and displays a clear transition
between the ergodic and localized behavior at a $\Delta J$ consistent
with the value estimated above, which seems
to become sharp in the thermodynamic limit $N\to\infty$.
Second, we study the disorder averaged entanglement entropy $\langle S\rangle$.
\begin{figure}\begin{center}
 \includegraphics[width=8cm]{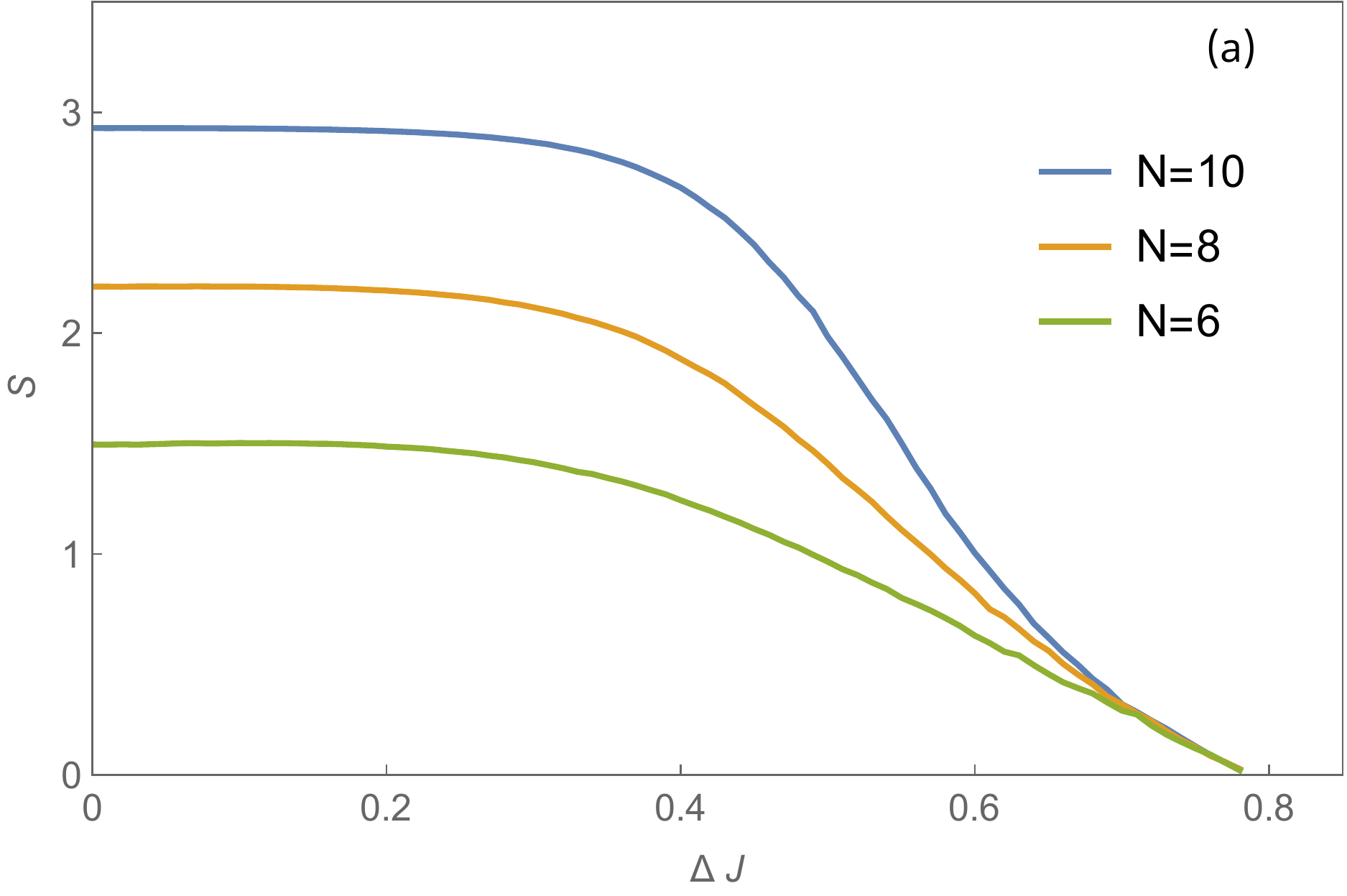}
  \includegraphics[width=8cm]{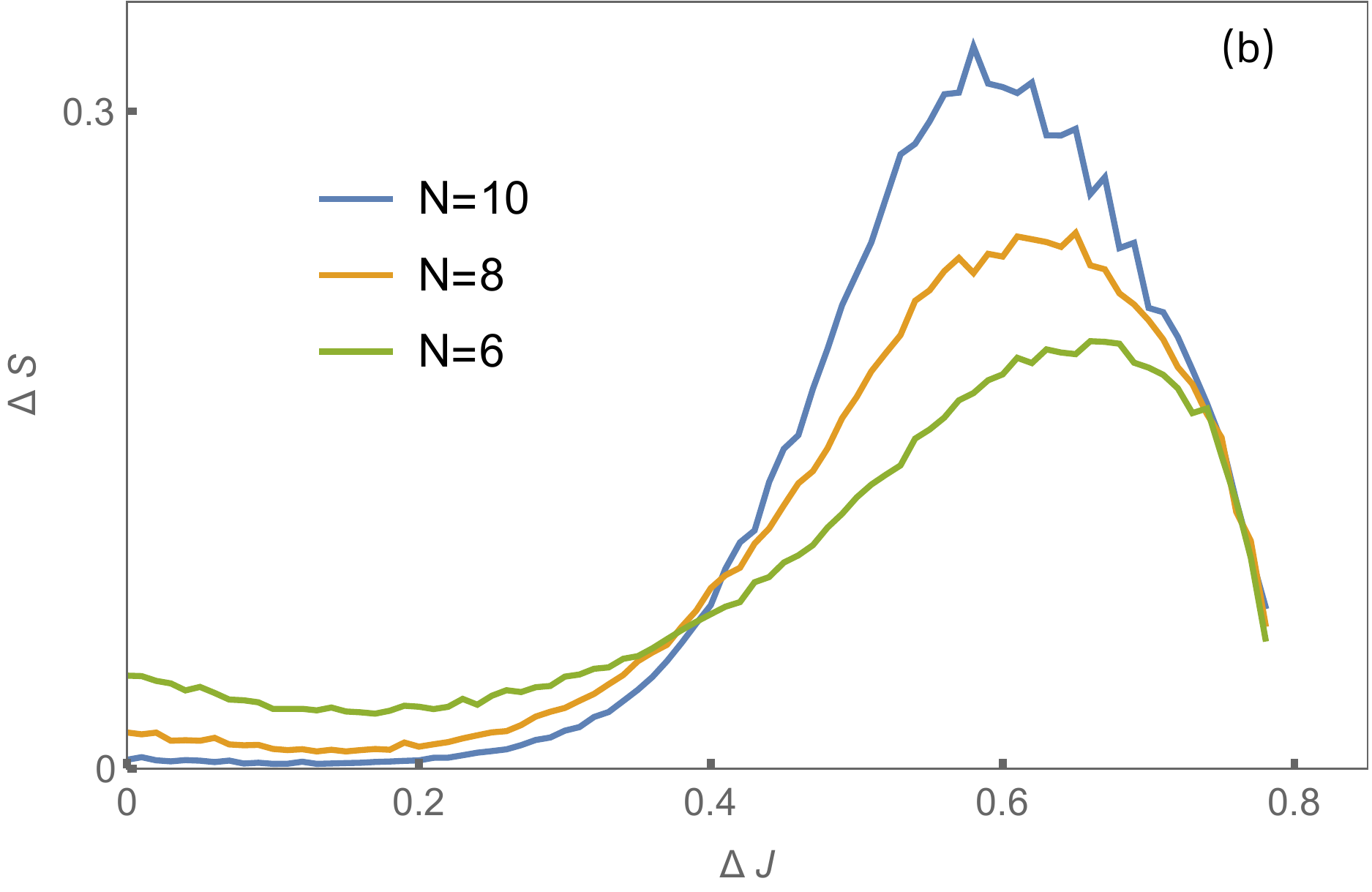}
\caption{Average of the entanglement entropy (a) of the symmetrically split system averaged with respect
to 1000 realizations and its variance (b). The upper curve for $S$ corresponds to $N=10$, the middle one to $N=8$ and the lower one to $N=6$. Fluctuations at the transition point
increase with the particle number.}
\label{entent}\end{center}
 \end{figure}
For the average we find a change from a volume-law ($\propto N$) behavior
to an area law ($\propto N^0$) in dependence of $\Delta J$, the variance shows a maximum and thereby a transition
again at the $\Delta J$ value just obtained, see Fig.\ \ref{entent}.
All these features are characteristic for a transition to many-body
localization \cite{Ponte}.

\section{Conclusions} The disordered spin chain with parameters satisfying the self-duality
condition is a rare example of a system in which absence of many-body
localization and adherence to the RMT predictions in the thermodynamic limit
have been analytically proven. We investigated
the break-up of these properties in the case of deviations from the self-dual situation.
A basic ingredient was the symmetry relation for the operators $\hat{U}$ and $\hat{W}_n$
with respect to the self-dual case. We explored its important consequences for the eigenphase density
which are in accordance with non-standard RMT ensembles. Furthermore, we studied $K_N(T)$ in dependence of $\Delta J$, $T$ and $N$ numerically and also explained our findings analytically.
 Finally, we can  identify a transition to MBL in the system.
This allows, first, to establish a connection between the Refs.\ \cite{Prosen0} and \cite{Prosen} by providing the time dependence of the spectral form factor.
Second, in view of Eq.\ (\ref{ffc}), the relationship between Refs.\ \cite{Prosen} and \cite{Chalker} becomes obvious, the RMT behavior shows up only in a narrow region in the
vicinity of the self-dual situation turning to a localized behavior of an interacting many-body system for increased $\Delta J$.

The methods developed here can be used to study many other quantities, as e.g.\
correlation functions and the entanglement entropy for spin chains which are in the focus of experimental and theoretical research.

\section{Acknowledgements}  We thank F. Hucht and M. Kieburg for fruitful discussions.

\appendix
\section{Perturbative Expansion of the form factor}
We now turn to the proof of the relation (\ref{bk}).
As one important input  we first determine the dimension of the subspaces labeled by $k$.
Defining projection operators onto that subspaces
\begin{eqnarray}
 \hat{Q}^{(0)}_\pm&=&\frac{1}{2T}(1\pm \hat{P}_R)\sum_{t=1}^T
 \left(\hat{P}_C\right)^t,\nonumber\\ \hat{Q}^{(k)}&=&\frac{1}{T}\sum_{t=1}^T
 {\rm e}^{i2\pi kt/T}\left(\hat{P}_C\right)^t,
\end{eqnarray}
we get for $T$ prime the dimensions of the subspaces by calculating the traces of the latter operators
\begin{eqnarray}
M_{0,\pm}&=&(2^T+2(T-1)\pm T 2^{(T+1)/2})/2T \nonumber\\
M_{k}&=& (2^T-2)/T.
\end{eqnarray}
In this context we used the relations
\begin{equation}
\mbox{Tr}\left(\hat{P}_C\right)^t = 2 \mbox{ for } t\neq T,   \qquad \mbox{Tr}\left(\hat{P}_C\right)^T = 2^T
\end{equation}
\begin{equation}
\mbox{Tr}\left(\hat{P}_R\left(\hat{P}_C\right)^t\right) = 2^{(T+1)/2} \mbox{ for all } t.
\end{equation}
Thereby we extend the analysis of Ref.\ \cite{Pineda}.

Furthermore, we need the relation
\begin{eqnarray}\label{eqbk}
 B_k&=&\frac{1}{2M_k}\left.\frac{\partial^2H(k,J)}{\partial J^2}\right|_{\Delta J=0}\nonumber\\
 \text{with}&&\hspace*{4mm}H(k,J)={\rm Tr}\widehat{\overline{W}}^{(k)}\left(\widehat{\overline{W}}^{(k)}\right)^\dagger
\end{eqnarray}
To show this we start from Eq.\ (\ref{fordis}), we insert
into that expression for $K_N(T)$ a power series of $\widehat{\overline{W}}$
\begin{equation}
\widehat{\overline{W}}=\widehat{\overline{W}}_0+\Delta J\widehat{\overline{W}}_1+\Delta J^2\widehat{\overline{W}}_2/2+\mathcal{O}(\Delta J^3).
\end{equation}
The zeroth order term yields the spectral form factor in the self-dual case $K_N^{\rm SD}(T)$, the
first order vanishes due to the symmetry relation (\ref{symmrelW}) and the second order reduces up to terms of order $\Delta J^3$ to
\begin{eqnarray}\label{eqlong}
 &&K_N(T)-K_N^{\rm SD}(T)\\&=&\frac{1}{2}N\Delta J^2{\rm Tr}\left\{\left[\hat{\mathcal{P}}\left(\widehat{\overline{W}}_0\otimes\widehat{\overline{W}}_0^*
 \right)\hat{\mathcal{P}}\right]^{N-1}\right.\nonumber\\ &&\left.\hat{\mathcal{P}}\left(\widehat{\overline{W}}_2\otimes\widehat{\overline{W}}_0^*+\widehat{\overline{W}}_0\otimes\widehat{\overline{W}}_2^*
 +2\widehat{\overline{W}}_1\otimes\widehat{\overline{W}}_1^*\right)\hat{\mathcal{P}}\right. \nonumber\\
 &+&\left.2\sum_{k=0}^{N-2}\left(\widehat{\overline{W}}_1\otimes \widehat{\overline{W}}_0^*+\widehat{\overline{W}}_0\otimes \widehat{\overline{W}}_1^*\right)
\left[\hat{\mathcal{P}}\left(\widehat{\overline{W}}_0\otimes \widehat{\overline{W}}_0^*\right)\hat{\mathcal{P}}\right]^k\right.\nonumber\\ &&\left.\left(\widehat{\overline{W}}_1\otimes \widehat{\overline{W}}_0^*+\widehat{\overline{W}}_0\otimes \widehat{\overline{W}}_1^*\right)
\left[\hat{\mathcal{P}}\left(\widehat{\overline{W}}_0\otimes \widehat{\overline{W}}_0^*\right)\hat{\mathcal{P}}\right]^{N-k-2}\right\}.\nonumber
\end{eqnarray}
We show first that the last summand in Eq.\ (\ref{eqlong})
 yields no contribution to the spectral form factor. For $N\gg1$ we can consider either $k$ or $N-k-2$ to be large
 implying that either $\left[\hat{\mathcal{P}}\left(\widehat{\overline{W}}_0\otimes \widehat{\overline{W}}_0^*\right)\hat{\mathcal{P}}\right]^{N-k-2}$ or
 $\left[\hat{\mathcal{P}}\left(\widehat{\overline{W}}_0\otimes \widehat
 {\overline{W}}_0^*\right)\hat{P}\right]^k$ acts as a projector onto the eigenvector to the eigenvalue one
 \begin{equation}\label{v0}
  v_0^{(k)}=\frac{1}{\sqrt{M_k}}\sum_{i=1}^{M_k}\left|\eta_i^{(k)}\right\rangle\left|\eta_i^{(k)}\right\rangle^*.
 \end{equation}
 This implies that the contribution given in the last two lines of Eq.\ (\ref{eqlong}) contains the factor
\begin{equation}\label{eq321}
\left\langle \vec{l}_1\vec{l}_2\left|\left(\widehat{\overline{W}}_1\otimes\widehat{\overline{W}}_0^*+\widehat{\overline{W}}_0\otimes
 \widehat{\overline{W}}_1^*\right)\right|v_0^{(k)}\right\rangle.
\end{equation}
with the basis states $\left|\vec{l}_1\right\rangle=\left|l_{11},\ldots,l_{1T}\right\rangle$ and $\left|\vec{l}_2\right\rangle=\left|l_{21},\ldots,l_{2T}\right\rangle$
and $l_{ij}\in\left\{-1,1\right\}$ that fulfill
the condition
\begin{equation}\label{euall}
\sum_{t=1}^Tl_{1t}=\sum_{t=1}^Tl_{2t}=n
\end{equation}
induced by $\hat{\mathcal{P}}$.
Using the special form of $v_0^{(k)}$ for $k\neq0$ Eq.\ (\ref{eq321}) can be rewritten as
\begin{eqnarray}\label{v0umf}
&&\frac{1}{\sqrt{M_k}}\left\langle \vec{l}_1\vec{l}_2\left|\left(\widehat{\overline{W}}_1\otimes\widehat{\overline{W}}_0^*+\widehat{\overline{W}}_0\otimes
 \widehat{\overline{W}}_1^*\right)\right|\sum_{i=1}^{M_k}\left|\eta_i^{(k)}\right>\left|\eta_i^{(k)*}\right>\right\rangle\nonumber
\\&&=\frac{1}{\sqrt{M_k}T}\sum_{i=1}^{M_k}\sum_{t,t'=1}^T \left\langle \vec{l}_1\vec{l}_2\left|\left(\widehat{\overline{W}}_1\otimes\widehat{\overline{W}}_0^*
+\widehat{\overline{W}}_0\otimes
 \widehat{\overline{W}}_1^*\right)\right|\nonumber\right.\\&&\left.{\rm e}^{2\pi i(t-t')k/T}\left(\hat{P}_C\right)^t\left|\vec{\sigma}^{(i)}\right>\left(\hat{P}_C\right)^{t'}\left|\vec{\sigma}^{(i)}\right>
 \right\rangle
\end{eqnarray}
The matrix elements of $\widehat{\overline{W}}_0$ in the basis $\left|\vec{\sigma}\right\rangle=\left|\sigma_1,\ldots,\sigma_T\right\rangle$ are given by
 \begin{eqnarray}\label{matrel}
  \left\langle\vec{\sigma}'\left|\widehat{\overline{W}}_0\right|\vec{\sigma}\right\rangle&=&\exp\left[-iJ\sum_{\tau=1}^T{\hat{\sigma}'}_\tau\hat{\sigma}_\tau-ih_z\sum_{\tau=1}^T
  \hat{\sigma}_\tau\right]\nonumber\\&&\prod_{\tau=1}^TR_{\sigma_\tau\sigma_{\tau+1}}(b_x),
 \end{eqnarray}
with $h_z$ being the disorder averaged magnetic field along the $z$-axis. The corresponding matrix elements of $\widehat{\overline{W}}_1$ are obtained from the last expression by taking the
first derivative with respect to $J$.
Eq.\ (\ref{eq321}) can be rewritten by using Eq.\ (\ref{matrel}) as
\begin{widetext}
\begin{eqnarray}
&&\frac{1}{\sqrt{M_k}T}\frac{\partial}{\partial J}\sum_{i=1}^{M_k}\sum_{t,t'=1}^T\exp\left[-iJ\sum_{\tau=1}^T\left(l_{1\tau}\sigma_{\tau+t}^{(i)}-l_{2\tau}\sigma_{\tau+t'}^{(i)}\right)\right]
{\rm e}^{2\pi i(t-t')k/T}\prod_{\tau=1}^T
 R_{l_{1\tau}l_{1\tau+1}}\left(b_x\right)R^*_{l_{2\tau}l_{2\tau+1}}\left(b_x\right)\\
 &=&\frac{1}{\sqrt{M_k}T}\frac{\partial}{\partial J}\sum_{i=1}^{M_k}\sum_{t,t'=1}^T\exp\left[-iJ\sum_{\tau=1}^T\left(l_{1\tau-t}-l_{2\tau-t'}\right)\left(\sigma_{\tau}^{(i)}-\sigma_{T}^{(i)}\right)\right]
{\rm e}^{2\pi i(t-t')k/T}\prod_{\tau=1}^T
 R_{l_{1\tau}l_{1\tau+1}}\left(b_x\right)R^*_{l_{2\tau}l_{2\tau+1}}\left(b_x\right)\nonumber,
\end{eqnarray}
\end{widetext}
where the last equality follows by the relation (\ref{euall}). To obtain the corresponding contribution to ${\rm Tr}\hat{\mathcal{A}}^N$
the latter expression needs to be evaluated on the self dual line implying $J=\pi/4$. Due to the fact that $\left(l_{1\tau-t}-l_{2\tau-t'}\right)$
as well as $\left(\sigma_{\tau}^{(i)}-\sigma_{T}^{(i)}\right)$ yield either $\pm2$ or $0$ it is easy to see that the latter expression is zero. We note in this context that for a
certain vector
$\left|\vec{\sigma}^{(i)}\right\rangle$
also the vector $-\left|\vec{\sigma}^{(i)}\right\rangle$ is contained in the $i$-summation given above
and second that the expression only involves first derivatives of $\cos(n\pi)$ with $n\in\mathds{N}$ which are zero. The calculation above applies to $k\neq 0$, the result carries over to
the terms involving $\hat{P}_R$ by replacing  $l_{1\tau}$ and $l_{2\tau}$ by $l_{1T-\tau}$ and $l_{2T-\tau}$, respectively.

We now turn to the first summand in Eq.\ (\ref{eqlong}).  
Using that $N$ is large, the $N-1$st power of $\left[\hat{\mathcal{P}}\left(\widehat{\overline{W}}_0\otimes\widehat{\overline{W}}_0^*
 \right)\hat{\mathcal{P}}\right]$ in the last equation acts as a projector onto its eigenvector
(\ref{v0})
to the eigenvalues of largest magnitude. The vectors $|\eta_i^{(k)}\rangle$ are defined in Eq.\ (\ref{eta}). Rewriting now first Eq.\ (\ref{eqlong}) and second Eq.\ (\ref{eqbk})
in terms of the matrix elements of the dual operator given
in Eq.\ (\ref{dualop}), relation (\ref{eqbk}) follows.

This relation can be used finally to obtain Eq. (\ref{bk}). Therefore $H(k,J)$ can be rewritten as
\begin{equation}
 H(k,J)=\frac{1}{T}\sum_{t=1}^T{\rm e}^{2\pi ikt/T}Z^{(t)}(J)
\end{equation}
with
\begin{equation}
Z^{(t)}(J)=\frac{1}{2^T}\sum_{\sigma_i,\sigma_i'=\pm1} \exp\left[-iJ\sum_{i=1}^T
\sigma_i\left(\sigma_i'-\sigma_{i+t}'\right)\right].
\end{equation}
The resulting sums in the last equation can be calculated for $T$ prime. This is best done by first taking the derivative in Eq.\ (\ref{eqbk}). Then we get
\begin{equation}
 \left.\frac{\partial^2 Z^{(t)}}{\partial J^2}\right|_{\Delta J=0}=\sum_{p,l=1}^T\mathcal{Z}^{(t)}(p,l)
\end{equation}
where for $p\neq \ell$
 \begin{multline}\label{mathz1}
\mathcal{Z}^{(t)}(p,\ell)= \sum_{\sigma'_i=\pm1}\left(\prod^T_{i=1, i\neq p, \ell} \cos \frac{\pi}{4}(\sigma'_i - \sigma'_{i+t}) \right) \\
\left(\sin\frac{\pi}{4} (\sigma'_p - \sigma'_{p+t})  \sin\frac{\pi}{4} (\sigma'_\ell- \sigma'_{\ell+t})\right) (\sigma'_\ell- \sigma'_{\ell+t}) (\sigma'_p - \sigma'_{p+t})
\end{multline}
and  for $p= \ell$:
\begin{equation}\label{mathz2}
\mathcal{Z}^{(t)}(p,p)= -\sum_{\sigma'_i=\pm1}\left(\prod^T_{i=1} \cos\frac{\pi}{4} (\sigma'_i - \sigma'_{i+t}) \right)  (\sigma'_p - \sigma'_{p+t})^2 .
\end{equation}
Note that for $t=0$ the above expressions vanish automatically due to the last two terms, and we  immediately have   $\left.\partial^2\mathcal{Z}^{(t=0)}(J)/\partial J^2\right|_{\Delta J=0}=0$.  For $t \neq 0$
we now treat the two cases (\ref{mathz1}) and (\ref{mathz2}) separately.
In case of (\ref{mathz2}) the product does not vanish only for   the two sequences
\[\sigma'_1 =\sigma'_1 = \dots =\sigma'_T =1, \qquad \sigma'_1 =\sigma'_1 = \dots =\sigma'_T =-1,\]
So we get  $\mathcal{Z}^{(t)}(p,p)=0$.

In case of (\ref{mathz1}) the sequences which contribute  must satisfy the following conditions:
\[ \sigma'_i =\sigma'_{i+t} \mbox{ for } i\neq \ell,p; \qquad \sigma'_\ell =- \sigma'_{\ell+t}, \sigma'_p= - \sigma'_{p+t}.  \]
It is straightforward to see that there are exactly two such sequences, each contributing $4$ into the sum. As a result we have
\[\mathcal{Z}^{(t)}(p,\ell)=8,  \mbox{ for }  \ell \neq p. \]
    Summing   up all factors together we obtain
\begin{equation}
\left.\frac{\partial^2Z^{(t)}}{\partial J^2}\right|_{\Delta J=0}=8T(T-1)
\end{equation}
Finally, after taking into account dimensions $M_k$ yields
\begin{equation}
B_{k}=\frac{8T(T-1)}{2 M_k T}\sum_{t=1}^{T-1}  e^{i2\pi kt/T} =\frac{-2T(T-1)}{2^{T-1} -1}.
\end{equation}
and the first relation in Eq.\ (\ref{bk}).
For the sectors $0\pm$ the function $H(k,J)$ defined in Eq.\ (\ref{eqbk})
can be written in the form
\begin{equation}
 H(0\pm,J)=H_1(J)\pm H_2(J)
\end{equation}
with
\begin{equation}\label{h2}
 H_1(J)=\frac{1}{2T}\sum_{t=1}^TZ^{(t)}(J),\hspace*{3mm}H_2(J)=\frac{1}{2T}\sum_{t=1}^TS^{(t)}(J)
\end{equation}
with
\begin{equation}
 S^{(t)}(J)=\frac{1}{2^T}\sum_{\sigma_i,\sigma_i'=\pm1} \exp\left[-iJ\sum_{i=1}^T
\sigma_i\left(\sigma_i'-\sigma_{T-i+t}'\right)\right].
\end{equation}
From the calculations above we get
\begin{equation}
 \frac{\partial^2 H_1(J)}{\partial J^2}\Big |_{J=\pi/4}= \frac{1}{2T}\sum_{t=0}^{T-1} \sum_{p,\ell =1}^T \mathcal{Z}^{(t)}(p,\ell)=4(T-1)^2,
\end{equation}
It thus remains to evaluate the right expression in Eq.\ (\ref{h2}).
Therefore we write
\begin{equation}
 \frac{\partial^2 H_2(J)}{\partial J^2}\Big |_{\Delta J=0}= \frac{1}{2T}\sum_{t=0}^{T-1} \sum_{p,\ell =1}^T \mathcal{S}^{(t)}(p,\ell),
\end{equation}
where for $p\neq \ell$
 \begin{eqnarray}
\mathcal{S}^{(t)}(p,\ell)&=& \sum_{\sigma'_i=\pm1}\left(\prod^T_{i=1, i\neq p, \ell} \cos \frac{\pi}{4}(\sigma'_i - \sigma'_{T-i+t}) \right)\nonumber\\&&
 \left(\sin\frac{\pi}{4} (\sigma'_p - \sigma'_{T-p+t})  \sin\frac{\pi}{4} (\sigma'_\ell- \sigma'_{T-\ell+t})\right) \nonumber\\ &&(\sigma'_\ell- \sigma'_{T-\ell+t})(\sigma'_p - \sigma'_{T-p+t})  \label{nondiag2}
\end{eqnarray}
and  for $p= \ell$:
\begin{equation}
\mathcal{S}^{(t)}(p,p)=- \sum_{\sigma'_i=\pm1}\left(\prod^T_{i=1} \cos\frac{\pi}{4} (\sigma'_i - \sigma'_{T-i+t}) \right)  (\sigma'_p - \sigma'_{T-p+t})^2 . \label{diag2}
\end{equation}
In Eq.\ (\ref{diag2}) the product does not vanish only if
\[ \sigma'_i =\sigma'_{T-i+t} \mbox{ for } i=1,\dots T.  \]
A simple analysis shows that there are $2^{(T+1)/2}$ such sequences with each one contributing $0$ into the sum.
Accordingly, we get  $\mathcal{S}^{(t)}(p,p)=0$.
In Eq.\ (\ref{nondiag2}) the sequences which contribute to the sum  must satisfy the following conditions:
\[ \sigma'_i =\sigma'_{T-i+t} \mbox{ for } i\neq \ell,p; \qquad \sigma'_\ell =- \sigma'_{T-\ell+t}, \sigma'_p= - \sigma'_{T-p+t}.  \]
It is straightforward to see that there are exactly  $2^{(T+1)/2}$ such sequences, each contributing $4$ into the sum. As a result we have
\[\mathcal{S}^{(t)}(p,\ell)=4 \cdot 2^{(T+1)/2},  \mbox{ for }  \ell \neq p. \]
    Summing   up all factors together we obtain
\begin{equation}
 \frac{\partial^2 H_2(J)}{\partial J^2}\Big |_{\Delta J=0} = 2(T-1) 2^{(T+1)/2}.
\end{equation}
After taking into account the dimensions $M_k$ we get
\begin{eqnarray}
B_{0\pm}&=&\frac{4(T-1)(T-1\pm 2^{(T-1)/2})}{2 M_k }\\ &=&\frac{4T(T-1)(T-1\pm 2^{(T-1)/2})}{2^T+2(T-1)\pm T 2^{(T+1)/2}}
=\frac{2T(T-1)}{\pm  2^{(T-1)/2}+1}.\nonumber
\end{eqnarray}
and thereby the right relation in (\ref{bk}).

\clearpage

\begin{thebibliography}{10}
\bibitem{Valid} We find good agreement with numerical results already for $\xi\approx 1$.
\end{thebibliography}
\end{document}